\documentclass[journal]{IEEEtran}
 \newtheorem{definition}{\bf Definition}

  \newtheorem{thm}{\bf Theorem}

 \newtheorem{lemma}{\bf Lemma}
  \newtheorem{prop}{\bf Proposition}
  \usepackage{dirtytalk}
\usepackage{amsfonts}
\usepackage{amssymb}
\usepackage{amsmath,graphicx}
\usepackage{bbm}
\usepackage{breqn}
\usepackage{subcaption}
\usepackage{diagbox}
\usepackage{cite}
\usepackage{cases}
\usepackage{tabu}
\usepackage{url}
\usepackage{multicol}
\usepackage{color}
\usepackage[bottom]{footmisc}
\usepackage{environ}
\usepackage{tikz}
 \usepackage{pgfplots}
\pgfplotsset{compat=newest}
\usetikzlibrary{plotmarks}
\usetikzlibrary{arrows.meta}
\usepgfplotslibrary{patchplots}
\usepackage{grffile}

\usepackage{enumerate}
\usepackage[]{algpseudocode}
\algtext*{EndWhile}
\algtext*{EndIf}
\algtext*{EndFor}
\usepackage{algorithm}

\newcommand*{\colorboxed}{}
\def\colorboxed#1#{%
  \colorboxedAux{#1}%
}
\newcommand*{\colorboxedAux}[3]{%
  \begingroup
    \colorlet{cb@saved}{.}%
    \color#1{#2}%
    \boxed{%
      \color{cb@saved}%
      #3%
    }%
  \endgroup
}

\newcommand{\argmin}{\operatornamewithlimits{argmin}}
\newcommand{\argmax}{\operatornamewithlimits{argmax}}

\title{Collaborative Beamforming Under Localization Errors: A Discrete Optimization Approach}
\author{Erfaun Noorani$^{\star}$, Yagiz Savas$^{\star}$,  Alec Koppel,  John Baras, Ufuk Topcu, and Brian M. Sadler
\thanks{ $^{\star}$ E. Noorani and Y. Savas contributed equally to this work.}\thanks{ This work is supported by the collaborative agreement ARL DCIST CRA W911NF-17-2-0181. E. Noorani is a Clark Doctoral Fellow at the Clark School of Engineering.} \thanks{
E. Noorani and J. Baras are with the Department of Electrical and Computer Engineering at the University of Maryland, MD, USA. Emails: \{enoorani,baras\}@umd.edu.}\thanks{
 Y. Savas and U. Topcu are with the Department of Aerospace Engineering at the University of Texas at Austin, TX, USA. Emails: \{yagiz.savas, utopcu\}@utexas.edu.}\thanks{
A. Koppel and B. M. Sadler are with the U.S. Army Research Laboratory, MD, USA. Emails: \{alec.e.koppel.civ, brian.m.sadler6.civ\}@mail.mil.}
}

\date{}
\begin{document}
\maketitle
\begin{abstract}
We consider a network of agents that locate themselves in an environment through sensor measurements and aim to transmit a message signal to a base station via collaborative beamforming. The agents' sensor measurements result in localization errors, which degrade the quality of service at the base station due to unknown phase offsets that arise in the agents' communication channels. Assuming that each agent's localization error follows a Gaussian distribution, we study the problem of forming a reliable communication link between the agents and the base station despite the localization errors. In particular, we formulate a discrete optimization problem to choose only a subset of agents to transmit the message signal so that the variance of the signal-to-noise ratio (SNR) received by the base station is minimized while the expected SNR exceeds a desired threshold. When the variances of the localization errors are below a certain threshold characterized in terms of the carrier frequency, we show that greedy algorithms can be used to globally minimize the variance of the received SNR. On the other hand, when some agents have localization errors with large variances, we show that the variance of the received SNR can be locally minimized by exploiting the supermodularity of the mean and variance of the received SNR. In numerical simulations, we demonstrate that the proposed algorithms have the potential to synthesize beamformers orders of magnitude faster than convex optimization-based approaches while achieving comparable performances using less number of agents. 
\end{abstract}

\section{Introduction}
Collaborative beamforming is a wireless communication technique in which a network of agents collectively transmit a common message signal to a base station \cite{visotsky1999optimum,sidiropoulos2006transmit, Ochiai2005}. Compared to single-agent transmission, collaborative beamforming has the potential to increase the range and rate of communication, to improve the directivity of the beam pattern, or to decrease the agents' individual power consumption while achieving the same quality of service (QoS) at the base station \cite{Barriac2004,Jayaprakasam2017, mudumbai2009distributed}. For example, in a network of $N$ agents, if each agent transmits the message signal with a fixed power, collaborative beamforming can lead to a factor of $N^2$ increase in the signal-to-noise ratio (SNR) received by the base station.

When the agents are distributed in an environment, which is the case in most multi-agent planning scenarios, e.g.,  \cite{torreno2017cooperative,Georgeff1983Communication, yuen}, realizing the full potential of collaborative beamforming proves itself to be a challenge due to the unknown phase offsets that arise in the agents' communication channels \cite{mudumbai}. The unknown phase offsets, which degrade the QoS at the base station, has mainly two sources: synchronization errors between the agents' local oscillators and the agents' localization errors. The former source has been extensively studied in the literature, and there are now a number of decentralized algorithms that can be used to mitigate the undesired effects of synchronization errors between the agents \cite{Barriac2004,mudumbai2007feasibility,brown2008time,shi2016extendable,alvarez2018distributed}. The latter source commonly arises in practice since the agents typically estimate their positions using sensor measurements resulting in an error associated with position \cite{fox2001particle,se2002mobile,hennes2012multi,artunedo}. The objective of this paper is to develop efficient beamforming methods that optimize the QoS at the base station despite the agents' localization errors.


From the analysis perspective, prior work focuses on understanding the effects of localization errors in collaborative beamforming. The agents' localization errors are associated with the topology of the network and translate to phasing errors in the transmission \cite{Ochiai2005}. The beam pattern characteristics for randomly generated network topologies is analyzed using the random array theory \cite{Ochiai2005,ahmed2009collaborative,huang2012collaborative,lo1964mathematical}. Specifically, the authors in \cite{ahmed2009collaborative} consider a setting in which each agent's location in the environment is sampled from the \textit{same} Gaussian distribution. They prove that, in this setting, the expected SNR received by the base station decays exponentially with a rate proportional to the variance of the Gaussian distribution. In \cite{mudumbai2007feasibility}, the authors show that, when the agents have fixed transmission powers and their phasing errors are \textit{identically} distributed, the expected SNR increases quadratically with the number of agents so long as the expected cosine of the phasing errors is close to one. 

In many multi-agent planning scenarios, the agents' position estimates follow \textit{non-identical} distributions, in which case the aforementioned results cannot be used to synthesize effective beamformers. Accordingly, in this paper, we consider a setting in which the agents' localization errors follow Gaussian distributions with potentially different mean and covariance. We derive the first- and second-order statistics of the received SNR as a function of the subset of agents that transmit the message signal. We then utilize the derived statistics to develop algorithms that include only a subset of agents in beamforming to optimize the QoS at the base station. 

From the algorithmic perspective, different approaches are proposed to synthesize beamformers that mitigate the undesired effects of phasing errors in the transmission. Ideally, with no localization and synchronization errors, the topology of the network may be used to derive the optimal beamformer, and we refer to this as the perfect channel state information (CSI) case. The work \cite{mudumbai} considers a case in which no CSI is available at the agents, i.e., the agents have no statistical information regarding their localization errors. The authors propose an iterative algorithm that maximizes the SNR at the base station by receiving feedback from the base station at each iteration. Subsequent work further investigates how the received feedback can be utilized to improve the reliability and security of communication \cite{song2011exploiting,tseng2014robust,Justin}. Although feedback-based approaches successfully improve the QoS at the base station, such approaches are iterative in nature. Therefore, their convergence to desired QoS levels may, in general, require a considerable number of iterations depending on network topology.
In the case of imperfect CSI, i.e., when a statistical information of the \textit{channel} is available at the agents, algorithms based on semi-definite programs (SDPs) are proposed to ensure that the received SNR is above a threshold with desired probability \cite{wang2010semidefinite,wang2011probabilistic,wang2014outage}. Similar conic optimization-based formulations are also common in the robust beamforming literature \cite{pascual2005robust,lorenz2005robust,luo2010semidefinite,ghavarol}. While SDP formulations provide a powerful method to improve the QoS without requiring feedback from the base station, they are computationally expensive and do not scale well with the number of agents.


In this paper, we approach the beamformer design problem from a discrete optimization perspective and develop three algorithms to choose a subset of agents to transmit the message signal to the base station. Given a network of agents with associated Gaussian localization errors, we seek a subset of agents to form a beam that achieves the desired QoS requirements without receiving feedback from the base station. To the best of our knowledge, this paper is the first one to employ discrete optimization techniques for mitigating the effects of localization errors in collaborative beamforming with provable performance guarantees. The main contributions of this paper are as follows:
\begin{itemize}
    \item First, under the assumption that each agent has a Gaussian localization error, we derive explicit forms of the expected value and the variance of the received SNR as a function of the agents that are included in beamforming. Using the derived expressions, we formulate a novel risk-sensitive discrete optimization problem: find a subset of agents to transmit the message signal such that the variance of the SNR at the base station is minimized while the expected SNR exceeds a desired threshold.
    \item Second, we propose two efficient sorting-based algorithms, Greedy and Double-Loop-Greedy (DLG), to solve the formulated discrete optimization problem and present sufficient conditions for their optimality. In particular, we show that the proposed algorithms return an optimal subset if the variance of the agents' localization errors is below a certain threshold which is characterized in terms of the carrier frequency. 
    \item Third, we prove that the expected value and the variance of the received SNR are supermodular set functions. Using this property, we develop a third algorithm, Difference-of-Submodular (DoS), which returns a subset that is \textit{locally} optimal for a certain relaxation of the formulated discrete optimization problem. The DoS algorithm utilizes the so-called submodular-supermodular procedure \cite{Narasimhan2005} as a subprocedure, and its local optimality guarantee is independent of the carrier frequency.
\end{itemize} 

The results presented in this paper show that, in the presence of localization errors, we may achieve the full potential of collaborative beamforming with minimum variability by including only a subset of the agents in beamforming. In particular, when the agents have small localization errors characterized in terms of the carrier frequency, we can globally minimize the variance of the SNR received by the base station using sorting-based algorithms. On the other hand, if some of the agents violate the small localization error condition, we can locally minimize the variance of the received SNR by exploiting the supermodularity of the mean and variance of the SNR. 


In numerical simulations, we compare the performance of the proposed algorithms with an SDP-based beamformer and demonstrate that all three algorithms, i.e., Greedy, DLG, and DoS, exhibit similar performances to that of the SDP-based beamformer while using less number of agents in beamforming. Moreover, for problem instances with large number of agents, Greedy and DLG algorithms compute the agent subset orders of magnitude faster than the SDP-based beamformer. 

\noindent\textbf{Related work:} 
A preliminary version of this paper appeared in \cite{noorani}, where we present the Greedy algorithm to solve the subset selection problem formulated in this paper. The major differences of this considerably extended version from the preliminary version are the following. First, we provide a numerical example illustrating the potential suboptimality of the Greedy algorithm and present the DLG algorithm, which improves the empirical performance over the Greedy algorithm. Second, we prove the supermodularity of the mean and variance of the received SNR as a function of the selected agent subsets and present the DoS algorithm to locally minimize the variance of the received SNR. Third, we provide numerical simulations to compare the performance of the Greedy, DLG, and DoS algorithms. Finally, we provide detailed proofs for all technical results. 

In addition to the aforementioned references, the subject of this paper is also related to the beamformer design when the agents have only local position information. Specifically, in \cite{vincent2007beamforming, papalexidis2007distributed}, the authors consider a setting in which the global location information is not available at the agents and design an antenna array that approximates the performance of a linear antenna array using only the information of exact inter-agent distances. Here, we consider a setting in which the statistics of the global location information is available at the agents. Hence, both the problem formulation and the proposed solution approaches are considerably different from the ones presented in \cite{vincent2007beamforming,papalexidis2007distributed}.

The idea of using only a subset of available agents in beamforming is previously investigated in the literature for various purposes. In \cite{ahmed2009collaborative,chen_side_lobe,sun2016node}, the authors choose a subset of sensor nodes to control the maximum sidelobe level. The work \cite{shulkind2018sensor} develops a discete-optimization based algorithm to design a sensor array for spatial sensing applications. Finally, the reference \cite{mehanna2013joint} studies the antenna selection problem in multicast beamforming. Unlike the above references, we consider the problem of achieving the desired SNR level at the base station with minimum variability despite localization errors and design discrete optimization-based algorithms that have provable performance guarantees.

\section{System Model} \label{System Model} 

We consider a group of $N$$\in$$\mathbb{N}$ agents that are distributed in an environment. Each agent is equipped with a single ideal isotropic antenna with a constant transmit power $P$$>$$0$. The agents' objective is to transmit a common message signal $m(t)$$\in$$\mathbb{R}$ to a base station equipped with a single antenna.

\subsection{Communication Channel}

Regarding the communication channel between the agents and the base station, we make the following assumptions.
\begin{enumerate}
    \item The transmitted signal $m(t)$ propagates in free space with no reflection or scattering.
    \item There is no mutual coupling effects between the agents' antennas.
    \item The local oscillators of all agents are time- and frequency-synchronized.
    \item The base station is located in the far-field region.
    \item The agent $i$$\in$$[N]$ transmits the signal $m(t)$ over a narrowband wireless channel $h_i$$\in$$\mathbb{C}$.
    \item All channels attenuate the signal $m(t)$ at the same level, i.e., $\lvert h_i\rvert$$=$$\lvert h_j\rvert$ for all $i,j$$\in$$[N]$.
\end{enumerate}

The assumption that the signal propagates in free space may hold in cluttered environments when the agents communicate with the base station at low VHF frequencies \cite{choi2017low,dagefu2015performance}. Similarly, mutual coupling effects may be avoided when the agents are sufficiently separated from each other. To achieve the frequency and time synchronizations, the agents may follow a separate short-range radio protocol \cite{mudumbai2007feasibility, mudumbai2009distributed}. Finally, the signal attenuation for all channels may be similar in scenarios in which the distance between the agents and the base station is significantly larger than inter-agent distances.

\subsection{Collaborative Transmission Model}
 
We consider a subset $\mathcal{S}$$\subseteq$$[N]$ of agents that collectively transmit the message signal $m(t)$ to the base station. All agents modulate $m(t)$ with the carrier signal $\operatorname{Re}\{e^{j2\pi f_c t}\}$, where $f_c$ is the carrier frequency. Each agent $i$$\in$$\mathcal{S}$ adjusts the phase of the transmission with the complex gain $w_i$$\in$$\mathbb{C}$ where $\lvert w_i\rvert$$=$$\sqrt{P}$, i.e., the transmit power is $P$. Then, the signal received by the base station is 
\begin{align*}
    y_{\mathcal{S}}(t)&:=\operatorname{Re}\Bigg\{e^{j2\pi f_c t}m(t)\sum_{i\in \mathcal{S}} w_i h_i \Bigg\}+n(t)
\end{align*}
where $n(t)$ is additive white Gaussian noise. Without loss of generality, we let $w_i$$=$$\sqrt{P}e^{j\delta_i}$ and $h_i$$=$$a_i e^{j\eta_i}$ for each $i$$\in$$[N]$. The angle $\delta_i$$\in$$[0,2\pi)$ denotes the phase of the gain $w_i$, and it is a design parameter. The magnitude $a_i$$>$$0$ and the phase $\eta_i$$\in$$[0,2\pi)$ characterize the channel $h_i$ between the base station and the agent $i$$\in$$[N]$. Recall that the agents' local oscillators are time-synchronized. Then, the phase offset $\eta_i$ of a signal at the base station relative to a signal transmitted by an agent located at $\vec{r}_i$$\in$$\mathbb{R}^3$ (in Cartesian coordinates) is \cite{mailloux1982phased}
\begin{align}\label{phase_offset_def}
   \eta_i = -\frac{2\pi f_c}{C} \langle \vec{r}_i, \vec{r}_c  \rangle.
\end{align}
In \eqref{phase_offset_def}, $\vec{r}_c$$\in$$\mathbb{R}^3$ is the unit vector pointing in the \textit{known} direction of the base station, $C$ is the speed of light, and $\langle \cdot,\cdot \rangle$ is the inner product of two vectors. 

We assume that the agents' local positions $\{\vec{r}_i : i$$\in$$[N]\}$ are not exactly known. In particular, for $i$$\in$$[N]$, we assume that $\vec{r}_i$$\sim$$\mathcal{N}({\boldsymbol{\mu}}_i,\Sigma_i)$ where ${\boldsymbol{\mu}_i}$$\in$$\mathbb{R}^3$ and $\Sigma_i$$\in$$\mathbb{R}^{3\times 3}$ are, respectively, the known mean and the known covariance of the Gaussian distribution. We note that the first and second order statistics of position estimates are typically easy to obtain in practice \cite{thrun2001robust,chen2003bayesian}. Finally, we assume that $\vec{r}_i$ and $\vec{r}_j$ are independent for $i,j$$\in$$[N]$ such that $i$$\neq$$j$. 

For a given subset $\mathcal{S}$$\subseteq$$[N]$ and the corresponding phase parameters $\delta_i$ for each $i$$\in$$\mathcal{S}$, let the \textit{array factor} be
\begin{align*}
    F(\mathcal{S},\delta):=\Bigg\lvert \sum_{i\in \mathcal{S}} e^{j (\delta_i+\eta_i)}\Bigg\rvert
\end{align*}
where $\delta$$:=$$[\delta_i | i $$\in$$ \mathcal{S}]$ is the vector of phase parameters. Assuming that $\lvert h_i\rvert$$=$$\lvert h_j\rvert$ for all $i,j$$\in$$[N]$, the magnitude of the array factor is proportional to the square root of the SNR received by the base station \cite{mudumbai}. Let the \textit{total phase} be $\Phi_i$$:=$$\delta_i+\eta_i$. The square of the the array factor yields the \textit{beamforming gain} $G(\mathcal{S},\delta)$ that is proportional to the received SNR and given by
\begin{align}\label{beamf_gain_first}
    G(\mathcal{S},\delta):= F^2(\mathcal{S},\delta)=\sum_{i\in \mathcal{S}}\sum_{j\in \mathcal{S}} \cos\Big(\Phi_i-\Phi_j\Big).
\end{align}
%

\section{Problem Statement} \label{Problem Statement} 
The beamforming gain $G(\mathcal{S},\delta)$ is a fundamental quantifier of the quality of a communication link with the base station as it is proportional to the received SNR. Hence, to establish a reliable communication link, we want the beamforming gain to be high with minimum variability. 

When the relative phase offsets $\eta_i$ are known, one can maximize $G(\mathcal{S},\delta)$ by selecting a pair $(\widetilde{\mathcal{S}},\widetilde{\delta})$ such that
\begin{align*}
    (\widetilde{\mathcal{S}},\widetilde{\delta})\in\argmax_{\substack{\mathcal{S}\subseteq[N],\\ \delta\in [0,2\pi)^N}}G(\mathcal{S},\delta).
\end{align*}
An optimal solution to the above optimization problem can be obtained by choosing $\widetilde{\mathcal{S}}$$=$$[N]$ and $\widetilde{\delta}_i$$=$$-\eta_i$ for all $i$$\in$$\widetilde{\mathcal{S}}$. To see this, recall that the total phase $\Phi_i$$=$$\delta_i+\eta_i$, and note that
\begin{align}
    G(\mathcal{S},\delta)=\sum_{i\in \mathcal{S}}\sum_{j\in \mathcal{S}} \cos\Big(\Phi_i-\Phi_j\Big)\leq N^2. \label{iff_upper_bound}
\end{align}
The upper bound in \eqref{iff_upper_bound} is attained if and only if $\mathcal{S}$$=$$[N]$ and $\Phi_i$$=$$\Phi_j$ for all $i,j$$\in$$\mathcal{S}$, i.e., the total phases are aligned. 

In this paper, we focus on a scenario in which $G(\mathcal{S},\delta)$ is a random variable since the agents' local positions $\{\vec{r}_i$$:$$i$$\in$$[N]\}$ are such that $\vec{r}_i$$\sim$$\mathcal{N}({\boldsymbol{\mu}}_i,\Sigma_i)$. In such a scenario, a reasonable objective might be to maximize the \textit{expected} beamforming gain by selecting a pair $(\overline{\mathcal{S}},\overline{\delta})$ such that
\begin{align}\label{stochastic_objective}
   (\overline{\mathcal{S}},\overline{\delta})\in\argmax_{\substack{\mathcal{S}\subseteq[N],\\ \delta\in [0,2\pi)^N}}\mathbb{E}\Big[G(\mathcal{S},\delta)\Big].
\end{align}

In Section \ref{property}, we show that the pair $(\mathcal{S},\delta)$$=$$([N],\hat{\delta})$, where $\hat{\delta}$$=$$[\hat{\delta}_i | i\in[N]]$ such that
\begin{align*}
     \hat{\delta}_i:=-\mathbb{E}[\eta_i] \ \ \text{for all}\ \  i\in[N],
\end{align*}
constitutes a solution to the problem in \eqref{stochastic_objective}. In other words, $\mathbb{E}[G(\mathcal{S},\delta)]$ is maximized by including all the agents in beamforming and aligning their total phases $\Phi_i$ \textit{in expectation}. 

Although including all the agents in beamforming maximizes the \textit{expected} beamforming gain, due to the random phase errors, this approach may actually decrease the probability with which the beamforming gain exceeds a certain threshold. To overcome this undesirable effect, we consider the variance of the beamforming gain as a risk measure and formulate a discrete optimization problem that includes only a subset of the agents in beamforming. In particular, we first fix the vector $\delta$ of phase parameters such that the expected beamforming gain is maximized, i.e., $\delta$$=$$\hat{\delta}$. Then, we aim to choose a subset $\mathcal{S}$$\subseteq$$[N]$ of agents that minimizes the variance of the beamforming gain while ensuring that the expected beamforming gain exceeds the desired threshold. The formal problem statement is as follows. 

\noindent \textbf{Problem 1: (Subset selection)} For a constant $\Gamma$$>$$0$, and the fixed vector of phase parameters $\delta$$=$$\hat{\delta}$, find $\mathcal{S}^{\star}$$\subseteq$$[N]$ such that 
\begin{subequations}
\begin{align}\label{opt_main_1}
       \mathcal{S}^{\star}\in\  &\argmin_{ \mathcal{S}\subseteq [N]}  \quad\ \  \mathrm{Var}\Big(G (\mathcal{S},\hat{\delta})\Big)\\ \label{opt_main_2}
        &\text{subject to:} \quad   \mathbb{E}\Big[G (\mathcal{S},\hat{\delta})\Big]\geq \Gamma.
\end{align}
\end{subequations}

Formulations that are similar to the subset selection problem are widely used in risk-sensitive optimization models \cite{levy1979approximating,markowitz2000mean}. By formulating a discrete optimization problem, we aim to design algorithms that improve the reliability of the communication link by minimizing the variability of received SNR and utilizing only a subset of the agents in beamforming. Finally, we remark that, in the subset selection problem, each agent $i$$\in$$[N]$ needs only its own position information, i.e., distribution of $\eta_i$, to set $\delta$$=$$\hat{\delta}$. Hence, in the considered setting, the agents adjust their phases in a distributed manner. 


\section{Statistical Properties of the Beamforming Gain }
\label{property}
In this section, we first derive the explicit form of the expected beamforming gain $\mathbb{E}[G(\mathcal{S},\delta)]$ and show that the pair $(\mathcal{S},\delta)$$=$$([N],\hat{\delta})$ is a solution to the problem in \eqref{stochastic_objective}. We then set $\delta$$=$$\hat{\delta}$ and derive the explicit form of $\mathrm{Var}(G(\mathcal{S},\hat{\delta}))$. The derived explicit forms are utilized to develop subset selection algorithms in the following sections.

Consider the definition of $G(\mathcal{S},\delta)$, given in \eqref{beamf_gain_first}, and recall that, for all $i$$\in$$[N]$, $\Phi_i$$=$$\eta_i$$+$$\delta_i$ where $\eta_i$$=$$2\pi f_c\ \langle \vec{r}_i, \vec{r}_c \rangle/ C$ and $\vec{r}_i$$\sim$$\mathcal{N}(\boldsymbol{\mu}_i,\Sigma_i)$. Then, for a given vector $\delta$$\in$$[0,2\pi)^N$ of phase parameters, we have $\Phi_i$$\sim$$\mathcal{N}(\theta_i,\gamma_i)$ where
\begin{flalign}\label{effective_var} \hspace{-0.1cm}
    \theta_i:= \frac{2\pi f_c}{C} \Big\langle \boldsymbol{\mu}_i, \vec{r}_c\Big\rangle+\delta_i \ \ \text{and} \ \  \gamma_i:= \frac{4\pi^2 f_c^2}{C^2} \Big\langle \vec{r}_c, \Sigma_i \vec{r}_c\Big\rangle. \raisetag{19pt}
\end{flalign}
We refer to $\gamma_i$ as \textit{the effective error variance} in the localization of the $i$th agent. Using the fact that $\Phi_i$$\sim$$\mathcal{N}(\theta_i,\gamma_i)$, we obtain the explicit form of $\mathbb{E}[G(\mathcal{S},\delta)]$ as follows.
{\setlength{\parindent}{0cm}\noindent
\begin{prop}\label{gaussian_power_prop22}
Let $v_i$$:=$$\exp(-\gamma_i)$. We have

\begin{align}\label{expected_closed_form_11}
    \mathbb{E}\Big[G(\mathcal{S},\delta)\Big]=\big\lvert \mathcal{S}\big\rvert +\sum_{i\in \mathcal{S}}\sum_{\substack{j\in \mathcal{S} \\ j\neq i}} \sqrt{v_i v_j} \cos(\theta_i-\theta_j).
\end{align}
\end{prop}}

The following result shows that we can maximize the expected beamforming gain by including all the agents in beamforming and aligning their phases in expectation.
{\setlength{\parindent}{0cm}\noindent
\begin{prop}\label{gaussian_power_prop33}
A pair $(\overline{\mathcal{S}},\overline{\delta})$ solves the problem in \eqref{stochastic_objective} if and only if $\overline{\mathcal{S}}$$=$$[N]$ and, for all $i,j$$\in$$[N]$, we have
\begin{align}
    \Big((\overline{\delta}_i+\mathbb{E}[\eta_i])- (\overline{\delta}_i+\mathbb{E}[\eta_i])\Big)\mod 2 \pi = 0.\label{mod_cond}
\end{align}
\end{prop}}

We provide proofs for the above propositions in Appendix \ref{appendix_A}. The condition $\overline{S}$$=$$[N]$ indicates that all the agents should be included in beamforming to maximize the expected beamforming gain. On the other hand, the condition in \eqref{mod_cond} implies that the agents' total phases should be aligned in expectation. Note that the vector $\hat{\delta}$, where $\hat{\delta}_i$$=$$-\mathbb{E}[\eta_i]$ for all $i$$\in$$[N]$, satisfies the condition in \eqref{mod_cond}. In the subset selection problem, we set $\delta$$=$$\hat{\delta}$ and aim to find a subset $\mathcal{S}$$\subseteq$$[N]$ that solves the risk-sensitive optimization problem given in \eqref{opt_main_1}-\eqref{opt_main_2}.

When $\delta$$=$$\hat{\delta}$, we have $\theta_i$$=$$\theta_j$ for all $i,j$$\in$$[N]$, implying that
\begin{align}\label{expected_closed_form}
        \mathbb{E}\Big[G(\mathcal{S},\hat{\delta})\Big]=\big\lvert \mathcal{S}\big\rvert +\sum_{i\in \mathcal{S}}\sum_{\substack{j\in \mathcal{S} \\ j\neq i}} \sqrt{v_i v_j}.
\end{align}
Next, we derive the variance of $G(\mathcal{S},\hat{\delta})$ as follows.
{\setlength{\parindent}{0cm}\noindent
\begin{prop}\label{gaussian_power_prop}
Let $v_i$$:=$$\exp(-\gamma_i)$. We have
%
\begin{align}
    \mathrm{Var}\Big(G(\mathcal{S},\hat{\delta})\Big)=&\sum_{i\in \mathcal{S}}\sum_{\substack{j\in \mathcal{S}\\ j\neq i}} \Big(1-v_iv_j\Big)^2\nonumber \\ \label{variance_closed_form}
    & +2\sum_{i\in \mathcal{S}}\sum_{\substack{j\in \mathcal{S}\\ j\neq i}}\sum_{\substack{k\in \mathcal{S}\\ k\neq i\\ k \neq j}} \Big(1-v_i\Big)^2\sqrt{v_j v_k}.
\end{align}
\end{prop}}

We provide a proof for the above result in Appendix \ref{appendix_A}. The proof exploits the equivalence $\mathbb{E}[\exp(tX)]$$=$$\exp(j\mu t-\sigma^2t^2/2)$ where $X$$\sim$$\mathcal{N}(\mu,\sigma)$ and the independence of $\vec{r}_i$ and $\vec{r}_j$ for $i$$\neq$$j$ to obtain the explicit form.


\section{Agent Selection Under Localization Errors} \label{results}
In this section, we propose three algorithms to solve the subset selection problem and analyze their optimality guarantees. Throughout this section, we assume that the problem in \eqref{opt_main_1}-\eqref{opt_main_2} has a feasible solution. For a given problem instance, the validity of this assumption can be easily verified by checking whether $\mathbb{E}[G([N],\hat{\delta})]$$\geq$$\Gamma$ due to the following result.
{\setlength{\parindent}{0cm}\noindent
\begin{prop}\label{exp_monotonocity}
For any $\mathcal{S}$$\subseteq$$\mathcal{S}'$$\subseteq$$[N]$, $\mathbb{E}[G (\mathcal{S},\hat{\delta})]$$\leq$$ \mathbb{E}[G (\mathcal{S}',\hat{\delta})]$.
\end{prop}}
The above result follows immediately from the fact that $\mathbb{E}[G(\mathcal{S},\hat{\delta})]$ is a sum of nonnegative terms; hence, adding an element to the subset can only increase the sum.
\subsection{Greedy Algorithm}
In this section, we consider a simple greedy algorithm to solve the subset selection problem and provide sufficient conditions for its optimality. The Greedy algorithm, shown in Algorithm \ref{greedy_algo}, first sorts the agents' effective error variances $\gamma_i$, defined in \eqref{effective_var}, in ascending order. We note that the sorting operation can be performed in $\mathcal{O}(N\log (N))$ for an array of length $N$ \cite{mergesort}. Initializing the output set $\mathcal{S}$ to the empty set, the algorithm then iteratively adds the agent with the next lowest effective error variance to the output set until the constraint $\mathbb{E}[G(\mathcal{S},\hat{\delta})]$$\geq$$\Gamma$ is satisfied.

  \begin{algorithm}[H]
\caption{Greedy}\label{greedy_algo}
\begin{algorithmic}[1]
\State \textbf{Input:} $\gamma_i$ for all $i$$\in$$[N]$, $\Gamma$$\in$$\mathbb{R}$.
\State Sort $\gamma_i$ such that $\gamma_{i_1}$$\leq$$\gamma_{i_2}$$\leq$$\ldots$$\leq$$\gamma_{i_N}$.
\State $\mathcal{S}$$:=$$\emptyset$, $k$$:=$$1$
\While {$\mathbb{E}[G(\mathcal{S},\hat{\delta})]$$<$$\Gamma$}
 \State $\mathcal{S}$$:=$$\mathcal{S}\cup \{i_k\}$, $k$$:=$$k+1$
\EndWhile
\State \textbf{return} $\mathcal{S}$.
\end{algorithmic}
\end{algorithm}

We now present sufficient conditions on the set $\{\gamma_i : i$$\in$$[N]\}$ for which the Greedy algorithm returns an optimal solution to the problem in \eqref{opt_main_1}-\eqref{opt_main_2}. Let \textit{the total effective error variance of a subset} $\mathcal{S}$$\subseteq$$[N]$ be measured by the function $V$$:$$2^{[N]}$$\rightarrow$$\mathbb{R}$ where $V(\mathcal{S})$$:=$$\sum_{i\in \mathcal{S}}\gamma_i$. Consider the problem of choosing a subset $\mathcal{S}'$$\subseteq$$[N]$ that satisfies the constraint in \eqref{opt_main_2} and has the minimum total effective error variance, i.e.,
\begin{subequations}
\begin{align}\label{opt_uncertainty_1}
       \mathcal{S}'\in\  &\arg \min_{ \mathcal{S}\subseteq [N]}  \quad\ V(\mathcal{S})\\ \label{opt_uncertainty_2}
        &\text{subject to:} \quad \ \mathbb{E}\Big[G (\mathcal{S},\hat{\delta})\Big]\geq \Gamma.
\end{align}
\end{subequations}
The next result, together with Proposition \ref{exp_monotonocity}, implies that the Greedy algorithm yields an optimal solution to the problem in \eqref{opt_uncertainty_1}-\eqref{opt_uncertainty_2}.
{\setlength{\parindent}{0cm}\noindent
\begin{prop}\label{expected_monotone_corollary}
For any $K$$\in$$\mathbb{N}$ such that $K$$\leq$$N$, we have
\begin{align*}
   \arg\min_{\substack{\mathcal{S}\subseteq [N]:\\  \lvert \mathcal{S}\rvert=K }} V(\mathcal{S}) =\arg\max_{\substack{\mathcal{S}\subseteq [N]:\\  \lvert \mathcal{S}\rvert=K} }\mathbb{E}\Big[G(\mathcal{S},\hat{\delta})\Big].
    \end{align*}
\end{prop}}

The above result follows from the fact that the derivative of the expected beamforming gain $\mathbb{E}[G(\mathcal{S},\hat{\delta})]$ with respect to $\gamma_i$, where $i$$\in$$\mathcal{S}$, is always negative. It can be shown that the problems in \eqref{opt_main_1}-\eqref{opt_main_2} and \eqref{opt_uncertainty_1}-\eqref{opt_uncertainty_2} are not equivalent in general. Hence, the greedy approach is, in general, not optimal to solve the subset selection problem. However, there are certain sufficient conditions, which are formalized below, under which such an approach becomes optimal.
{\setlength{\parindent}{0cm}\noindent
\begin{thm}\label{main_thm_1} For a given set $\{\gamma_i : i$$\in$$[N]\}$ of effective error variances, let $\gamma_{i_1}$$\leq$$\gamma_{i_2}$$\leq$$...$$\leq$$\gamma_{i_N}$ where $i_k$$\in$$[N]$. A solution to the problem in \eqref{opt_uncertainty_1}-\eqref{opt_uncertainty_2} is also a solution to the problem in \eqref{opt_main_1}-\eqref{opt_main_2} if either one of the following conditions hold:
\begin{description}
   \item[(C1)]  $\mathbb{E}[G(\mathcal{S},\hat{\delta})]$$\geq$$\Gamma$ where $\mathcal{S}$$=$$\{i_1,i_2\}$,
   \item[(C2)] $\gamma_{i_N}$$\leq$$0.83$.
\end{description}
\end{thm}}

We provide a proof for the above result in Appendix \ref{appendix_A}. The main idea in the proof is to show that the derivative of $\mathrm{Var}(G(\mathcal{S},\hat{\delta}))$ with respect to $\max_{i\in \mathcal{S}}\gamma_{i}$ is positive. Condition (C1) follows from the fact that, when $\lvert \mathcal{S} \rvert$$\leq$$2$, the derivative is always positive. Condition (C2) follows from the fact that, when $\gamma_{i_N}$$\leq$$0.83$, the derivative is positive regardless of the size of the set $\mathcal{S}$. For such $\gamma_{i_N}$, the subset with minimum total effective error variance is the one that minimizes the variance of the beamforming gain; hence, the problems in \eqref{opt_uncertainty_1}-\eqref{opt_uncertainty_2} and  \eqref{opt_main_1}-\eqref{opt_main_2} become equivalent when (C1) or (C2) holds.

Theorem \ref{main_thm_1} states that if all the agents have ``small" effective error variances, then the Greedy algorithm returns an optimal solution to the subset selection problem. In particular, it follows from Theorem \ref{main_thm_1} that a sufficient condition for optimality characterized by the carrier frequency is
\begin{align*}
  \max_{i\in [N]}\Big\langle \vec{r}_c, \Sigma_i \vec{r}_c\Big\rangle\leq \frac{0.83 C^2}{4\pi^2 f_c^2}.
\end{align*}

\begin{figure}[t!]
    \centering
%
%
\definecolor{mycolor1}{rgb}{0.00000,0.44700,0.74100}%
\begin{tikzpicture}

\begin{axis}[%
width=0.6\linewidth,
at={(1.011in,0.653in)},
scale only axis,
xmin=20,
xmax=200,
xlabel style={font=\color{white!15!black}},
xlabel={Carrier frequency $f_c$ (MHz)},
ymode=log,
ymin=0.01,
ymax=10,
yminorticks=true,
ylabel style={font=\color{white!15!black}},
ylabel={Maximum error variance $\sigma_{\max}^2\ (m^2)$},
axis background/.style={fill=white},
xmajorgrids,
ymajorgrids,
]
\addplot [color=mycolor1, line width=2.0pt, forget plot]
  table[row sep=crcr]{%
20	4.72388993542899\\
30	2.09950663796844\\
40	1.18097248385725\\
50	0.755822389668639\\
60	0.52487665949211\\
70	0.385623668198298\\
80	0.295243120964312\\
90	0.233278515329827\\
100	0.18895559741716\\
110	0.15616165075798\\
120	0.131219164873027\\
130	0.111808045808972\\
140	0.0964059170495744\\
150	0.0839802655187376\\
160	0.073810780241078\\
170	0.0653825596599184\\
180	0.0583196288324566\\
190	0.0523422707526767\\
200	0.0472388993542899\\
};
\end{axis}
\end{tikzpicture}%
    \caption{Maximum localization error variance $\sigma_{\max}^2$ allowed for the optimality of the greedy algorithm as a function of the carrier frequency $f_c$. Note that the localization error tolerance is relaxed at lower frequencies (longer wavelengths).}
    \label{comparison_plot}
\end{figure}
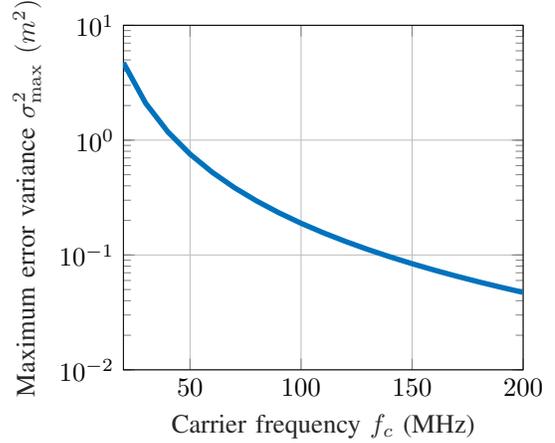

For example, suppose that $\Sigma_i$$=$$\sigma_i^2 I_{3\times 3}$, where $I_{3\times 3}$ is the identity matrix, and let $\sigma_{\max}^2$$:=$$\max_i \sigma_i^2$. Then, we have $\sigma_{\max}^2$$\leq$$ \frac{0.83 C^2}{4\pi^2 f_c^2}$ as the sufficient condition (C2). In Figure \ref{comparison_plot}, we graphically illustrate the trade-off between the carrier frequency $f_c$ and the maximum variance $\sigma_{\max}^2$ under which the Greedy algorithm is optimal. Note that as $f_c$ increases (resulting in shorter wavelength), condition (C2) requires smaller position error variance, whereas longer wavelengths increase the error tolerance. For example, at lower VHF frequencies, e.g., $f_c$$=$$40$ MHz for which the effective wavelength is $\lambda_c$$=$$C/f_c$$\approx$$7.5$ meters, the agents are allowed to have localization error variance up to 1 square meter. Hence, for this frequency range, the position error tolerance can easily be achieved with existing localization algorithms \cite{fox2001particle,lavalle2006planning}.

\subsection{Suboptimality of the Greedy Algorithm}

In this section, we illustrate with a numerical example that the Greedy algorithm may fail to return an optimal solution to the subset selection problem instances that violate the sufficient conditions presented in Theorem \ref{main_thm_1}.

\begin{figure}[t!]
\centering
%
%
\definecolor{mycolor1}{rgb}{0.00000,0.44700,0.74100}%
\definecolor{mycolor2}{rgb}{0.85000,0.32500,0.09800}%
\definecolor{mycolor3}{rgb}{0.92900,0.69400,0.12500}%
\begin{tikzpicture}

\begin{axis}[%
width=0.6\linewidth,
at={(1.011in,0.642in)},
scale only axis,
xmin=1,
xmax=4,
xtick={1, 2, 3, 4},
xlabel style={font=\color{white!15!black}},
xlabel={Subset number (k)},
ymin=3,
ymax=9,
ytick={1, 2, 3, 4, 5, 6, 7, 8,9},
ylabel style={font=\color{white!15!black}},
ylabel={Function value},
axis background/.style={fill=white},
xmajorgrids,
ymajorgrids,
legend style={at={(1,0.58)}, draw=white!15!black}
]
\addplot [color=mycolor1, densely dotted, line width=2.0pt, mark = o, mark options ={solid}]
  table[row sep=crcr]{%
1	6.97126370781247\\
2	7.24106364968566\\
3	6.82097079158534\\
4	6.76294479196693\\
};
\addlegendentry{$\mathrm{Var}(G(S_k, \hat{\delta}))$}

\addplot [color=mycolor2, dashed,line width=2.0pt, mark=o, mark options={solid}]
  table[row sep=crcr]{%
1	4.90902614397391\\
2	4.4690924701552\\
3	3.53640935136244\\
4	3.48884917947108\\
};
\addlegendentry{$\mathbb{E}[G(S_k,\hat{\delta})]$}

\addplot [color=mycolor3, mark = o, mark options = {solid}, line width=2.0pt]
  table[row sep=crcr]{%
1	4\\
2	6\\
3	8.4\\
4	8.6\\
};
\addlegendentry{$V(S_k)$}

\end{axis}
\end{tikzpicture}%
\caption{An illustration of the numerical example showing that the Greedy algorithm may return a suboptimal subset. The variance of the beamforming gain is minimized by the subset which has the maximum total effective error variance.  } \label{counter-example}
\end{figure}
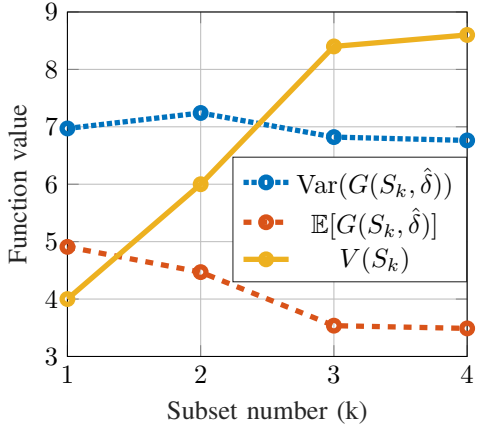

Let the total number of agents be $N$$=$$4$, and the expected gain threshold be $\Gamma$$=$$3.3$. Furthermore, let the ordered set $(\gamma_1,\gamma_2, \gamma_3, \gamma_4)$ of effective error variances be $(0.4,0.6,3,5)$. It can be shown by direct calculations that this problem instance violates the sufficient conditions presented in Theorem \ref{main_thm_1}. Moreover, using Proposition \ref{expected_monotone_corollary} and the monotonocity of $\mathrm{Var}(G(\mathcal{S},\hat{\delta}))$ in $\mathcal{S}$$\subseteq$$[N]$, it can be shown that $\lvert \mathcal{S}^{\star}\rvert$$=$$3$. Let $\mathcal{S}_1$$:=$$\{1,2,3\}$, $\mathcal{S}_2$$:=$$\{1,2,4\}$, $\mathcal{S}_3$$:=$$\{1,3,4\}$, and $\mathcal{S}_4$$:=$$\{2,3,4\}$ be all possible subsets of $\{1, 2, 3, 4\}$ containing three elements. In Figure \ref{counter-example}, we provide the expected value and the variance of the beamforming gain $G(\mathcal{S}_k,\hat{\delta})$ as well as the total effective error $V(\mathcal{S}_k)$ for each $\mathcal{S}_k$. All subsets $\mathcal{S}_k$ satisfy $\mathbb{E}[G(\mathcal{S}_k,\hat{\delta})]$$\geq$$\Gamma$. Observe that the optimal solution for this problem instance is the subset $\mathcal{S}_4$ which has the maximum total effective error variance $V(\mathcal{S}_4)$ instead of the minimum one $V(\mathcal{S}_1)$. Therefore, for this problem instance, the Greedy algorithm does not yield the optimal solution to the subset selection problem.

The above numerical example illustrates a counterintuitive fact: for some problem instances, the optimal subset consists of the agents with \textit{the highest effective localization error variances}. This phenomenon arises due to the constructive/destructive interference of the sinusoids in \eqref{beamf_gain_first}. Next, we modify the Greedy algorithm to account for the fact that for some instances the agents with the highest effective localization error variances may provide an optimal solution to the subset selection problem.


\subsection{Double-Loop-Greedy for Improved Empirical Performance}
Inspired by the numerical example given in the previous section, as the second approach to solve the subset selection problem, we propose the Double-Loop-Greedy (DLG) algorithm shown in Algorithm \ref{Double_loop_greedy_algo}. Similar to the Greedy algorithm, the DLG algorithm first sorts the agents' effective error variances $\gamma_i$ in ascending order.
It then initializes two sets, namely, $\mathcal{S}_1$ and $\mathcal{S}_2$, to the empty set. Starting from the agent with the lowest effective error variance, at each iteration, the agent with the next \textit{lowest} effective error variance is iteratively added to the set $\mathcal{S}_1$ until the constraint $\mathbb{E}[G(\mathcal{S}_1,\hat{\delta})]$$>$$\Gamma$ is satisfied. Note that $\mathcal{S}_1$ is the same as the Greedy algorithm. Similarly, starting from the agent with the highest effective error variance, at each iteration, the agent with the next \textit{highest} effective error variance is iteratively added to the set $\mathcal{S}_2$ until the constraint $\mathbb{E}[G(\mathcal{S}_2,\hat{\delta})]$$>$$\Gamma$ is satisfied. Finally, the DLG algorithm compares the variance of the beamforming gain for $\mathcal{S}_1$ and $\mathcal{S}_2$, and outputs the one with smaller value. We note that the time complexity of the DLG algorithm is the same as the time complexity of the Greedy algorithm.

\begin{algorithm} [t]
\caption{Double-Loop-Greedy (DLG)}\label{Double_loop_greedy_algo}
\begin{algorithmic}[1]
\State \textbf{Input:} $\gamma_i$ for all $i$$\in$$[N]$, $\Gamma$$\in$$\mathbb{R}$.
\State Sort $\gamma_i$ such that $\gamma_{i_1}$$\leq$$\gamma_{i_2}$$\leq$$\ldots$$\leq$$\gamma_{i_N}$.
\State $\mathcal{S}_1$$:=$$\emptyset$, $\mathcal{S}_2$$:=$$\emptyset$, $k$$:=$$1$, $l$$:=$$N$
\While {$\mathbb{E}[G(\mathcal{S}_1,\hat{\delta})]$$<$$\Gamma$}
 \State $\mathcal{S}_1$$:=$$\mathcal{S}_1\cup \{i_k\}$, $k$$:=$$k+1$
\EndWhile
\While {$\mathbb{E}[G(\mathcal{S}_2,\hat{\delta})]$$<$$\Gamma$}
 \State $\mathcal{S}_2$$:=$$\mathcal{S}_2\cup \{i_l\}$, $l$$:=$$l-1$
\EndWhile
\If{$\mathrm{Var}(G(\mathcal{S}_1,\hat{\delta}))$$<$$\mathrm{Var}(G(\mathcal{S}_2,\hat{\delta}))$} $\mathcal{S}:=\mathcal{S}_1$
\Else $\ \mathcal{S}:=\mathcal{S}_2$
\EndIf
\State \textbf{return} $\mathcal{S}$.
\end{algorithmic}
\end{algorithm}

\noindent {\bf Optimality of the DLG algorithm:}
For a given problem instance, the subset $\mathcal{S}$$\subseteq$$[N]$ returned by the DLG algorithm is guaranteed to satisfy $\mathrm{Var}(G(\mathcal{S},\hat{\delta}))$$\leq$$\mathrm{Var}(G(\mathcal{S}',\hat{\delta}))$ where $\mathcal{S}'$$\subseteq$$[N]$ is the subset returned by the Greedy algorithm. Hence, the DLG algorithm also provides an optimal solution to the problem in \eqref{opt_uncertainty_1}-\eqref{opt_uncertainty_2} under the sufficient conditions stated in Theorem \ref{main_thm_1}. Moreover, as can be seen from the numerical example given in the previous section, the DLG algorithm may also return an optimal solution to instances on which the Greedy algorithm performs poorly.

\subsection{Difference-of-Submodular (DoS) Algorithm} \label{supermodularity}
Both the Greedy and DLG algorithms are guaranteed to return optimal solutions to the subset selection problem \textit{under the sufficient conditions} stated in Theorem \ref{main_thm_1}. In this section, we propose a third approach to solve the subset selection problem, which \textit{always} returns a \textit{locally} optimal solution to a certain relaxation of the subset selection problem. Although the proposed third approach is computationally more demanding, its local optimality guarantee is independent of the carrier frequency unlike the greedy approaches.

Before presenting the proposed approach, i.e., Difference-of-Submodular (DoS) algorithm, we first provide a definition of submodularity and show that both  $\mathbb{E}[G (\mathcal{S},\hat{\delta})]$ and $\mathrm{Var}(G (\mathcal{S},\hat{\delta}))$ are supermodular set functions.
{\setlength{\parindent}{0cm}\noindent
\begin{definition} A set function $f$$:$$2^{\Omega}$$\rightarrow$$\mathbb{R}$ is submodular if for every $X, Y$$\subseteq$$\Omega$ with $X$$\subseteq$$Y$ and every $e$$\in$$\Omega\backslash Y$, we have $f(X$$\cup$$\{e\})- f(X) \geq f(Y\cup\{e\})- f(Y)$.
\end{definition}}
A set function $f$$:$$2^{\Omega}$$\rightarrow$$\mathbb{R}$ is said to be \textit{supermodular} if the set function $-f$ is submodular.
{\setlength{\parindent}{0cm}\noindent
\begin{thm}
\label{supermodularity_theorem}
Both $\mathbb{E}[G (\mathcal{S},\hat{\delta})]$ and $\mathrm{Var}(G (\mathcal{S},\hat{\delta}))$ are supermodular set functions.
\end{thm}}

A proof of Theorem \ref{supermodularity_theorem} is provided in Appendix \ref{appendix_A}. Next, we formalize the notion of local optimality for discrete optimization problems and introduce the DoS algorithm which utilizes the results of \cite{Narasimhan2005} as subprocedures.
{\setlength{\parindent}{0cm}\noindent
\begin{definition}
\cite{Narasimhan2005} For a set function $\phi$$:$$2^{\Omega}$$\rightarrow$$\mathbb{R}$, a sequence $\{S_t\subseteq \Omega : t\in \mathbb{N}\}$ is said to converge to a local minimum if there exists a constant $M$$\in$$\mathbb{N}$ such that $\phi(S_m)$$=$$\phi(S_n)$ for all $m,n$$\geq$$M$, and for any $k$$\in$$\mathbb{N}$, $\phi(S_k)$$\leq$$\phi(S_l)$ for all $l$$\leq$$k$.
\end{definition}}

Let $f$$:$$2^{\Omega}$$\rightarrow$$\mathbb{R}$ and $g$$:$$2^{\Omega}$$\rightarrow$$\mathbb{R}$ be submodular set functions. In \cite{Narasimhan2005}, the authors present an algorithm, called Submodular-Supermodular-Procedure (SSP), that returns a local optimal solution to the following problem
\begin{align}\label{opt_bar_bar}
       \min_{ S\subseteq \Omega}  \quad\ f(S) -  g(S).
\end{align}

The DoS algorithm, shown in Algorithm \ref{DS_algo}, utilizes the SSP as a subprocedure to return a locally optimal solution to a certain relaxation of the subset selection problem. In particular, it takes two parameters $\lambda_0$$>$$0$ and $\alpha$$>$$1$ as inputs as well as the agents' effective localization error variances $\gamma_i$ and the expected gain threshold $\Gamma$. At the $k$th iteration, where $k$$\in$$\mathbb{N}$, using the SSP as a subprocedure, the DoS algorithm finds a locally optimal solution to the following problem
\begin{align}\label{opt_bar}
       \min_{ \mathcal{S}\subseteq [N]}  \quad\ \mathrm{Var}\Big(G(\mathcal{S},\hat{\delta})\Big) - \lambda_k  \mathbb{E}\Big[G(\mathcal{S},\hat{\delta})\Big]
\end{align}
where $\lambda_k$ is iteratively defined as $\lambda_k$$=$$\alpha \lambda_{k-1}$. The algorithm terminates when the solution returned by the SSP satisfies $\mathbb{E}[G(\mathcal{S},\hat{\delta})]$$\geq$$\Gamma$.

\begin{algorithm} [t!]
\caption{Difference-of-Submodular (DoS)}\label{DS_algo}
\begin{algorithmic}[1]
\State \textbf{Input:}  $\gamma_i$ for all $i$$\in$$[N]$, $\Gamma$$\in$$\mathbb{R}$, $\lambda_0$$>$$0$, $\alpha$$>$$1$.
\State $\mathcal{S}$$:=$$\emptyset$, $k$$:=$$0$.
\While{$\mathbb{E}[G (\mathcal{S},\hat{\delta})]$$<$$\Gamma$}
\State $f(\cdot)$$:=$$-\lambda_k\mathbb{E}[G (\cdot,\hat{\delta})]$, $g(\cdot)$$:=$$-\mathrm{Var}(G (\cdot,\hat{\delta}))$
\State $\mathcal{S} $$:=$$ \text{SSP} (f (\cdot), g(\cdot) )$
\State $k$$:=$$k+1$, $\lambda_k$$:=$$\alpha \lambda_{k-1}$.
\EndWhile
\State \textbf{return} $\mathcal{S}$.
\end{algorithmic}
\end{algorithm}

 {\noindent \bf Convergence of the DoS algorithm:} For the DoS algorithm to terminate, the subprocedure SSP should output a subset $\mathcal{S}$$\subseteq$$[N]$ such that $\mathbb{E}[G (\mathcal{S},\hat{\delta})]$$\geq$$\Gamma$. At the $k$th iteration, the SSP finds a locally optimal solution to the problem in \eqref{opt_bar} by computing successive modular approximations of the function $\mathrm{Var}(G(\mathcal{S},\hat{\delta}))$ and finding a globally optimal solution to each of the resulting approximation problems. Since $\lambda_0$$>$$0$ and $\alpha$$>$$1$, the parameter $\lambda_k$ increases at each iteration. Hence, in terms of the objective value, the globally optimal solution of the approximate problems become closer to the globally optimal solution of $\max_{\mathcal{S}}\mathbb{E}[G (\mathcal{S},\hat{\delta})]$, which is $\mathcal{S}$$=$$[N]$.
Since we assumed at the beginning that there exists a feasible solution to the subset selection problem, the DoS algorithm is guaranteed to terminate for some finite $k$$\in$$\mathbb{N}$.

{\bf \noindent Optimality of the DoS algorithm:} As mentioned earlier, at each iteration, the DoS algorithm computes a locally optimal solution to the problem in \eqref{opt_bar}. Hence, the subset returned by the DoS algorithm is a locally optimal solution to the following relaxation of the subset selection problem
\begin{align}\label{opt_bar1}
       \min_{ \mathcal{S}\subseteq [N]}  \quad\ \mathrm{Var}\Big(G(\mathcal{S},\hat{\delta})\Big) - \lambda_{k^{\star}}  \mathbb{E}\Big[G(\mathcal{S},\hat{\delta})\Big]
\end{align}
where $k^{\star}$ is the number of iterations until the convergence of the DoS algorithm. We also note that the above problem formulation is sometimes referred to as a ``regularized version" of the original constrained optimization problem \cite{boyd2004convex}. Such regularization approaches to solve the original problems are also common in portfolio management \cite{neu2017unified} and reinforcement learning \cite{steinbach2001,koppel2017policy,koppel2018nonparametric}, among many others.

\section{Numerical Simulations} \label{Numerical Simulation}
We present numerical simulation results that demonstrate the performance of the proposed algorithms on randomly generated instances of the subset selection problem. All computations are run on a 3.1-GHz desktop with 32 GB RAM using the toolbox \cite{SFO_toolbox} for the implementation of the SSP (step 5 in the DoS algorithm).

\subsection{Suboptimality Ratio on Small-Scale Instances  }

We compare the suboptimality of the proposed algorithms as a function of three problem parameters: the total number $N$ of agents, the maximum localization error $\gamma_{\max}$$:=$$\min \{\gamma : \gamma$$\geq$$\gamma_i, \text{for all} \ i$$\in$$[N]\}$, and the expected beamforming gain threshold $\Gamma$$=$$\beta\Gamma_{\max}$ where $0$$<$$\beta$$\leq$$1$ and $\Gamma_{\max}$$:=$$\mathbb{E}[G([N],\hat{\delta})]$ is the maximum expected beamforming gain that can be achieved by the agents.

For a given problem instance, we measure the performance of an algorithm by the suboptimality ratio (SR) of its output. Specifically, let $\mathcal{S}^{\star}$ be an optimal solution to the given problem instance \eqref{opt_main_1}-\eqref{opt_main_2}, which, for small $N$, can be computed by considering all subsets $\mathcal{S}$$\subseteq$$[N]$. Moreover, let $\overline{\mathcal{S}}$ be the (possibly suboptimal) output of a given algorithm. We define the SR of the algorithm on the given instance as
\begin{align*}
    \text{SR}:= \frac{\mathrm{Var}\Big(G(\overline{\mathcal{S}},\hat{\delta})\Big)}{\mathrm{Var}\Big(G(\mathcal{S}^{\star},\hat{\delta})\Big)}.
\end{align*}
All proposed algorithms, i.e., Greedy, DLG, and DoS, have $\text{SR}$$\geq$$1$ since their output $\overline{\mathcal{S}}$ satisfies $\mathbb{E}[G(\overline{\mathcal{S}},\hat{\delta})]$$\geq$$\Gamma$.
\begin{figure*}[h]
    \centering
    \minipage[t]{1\linewidth}
	\begin{subfigure}[t]{0.34\linewidth}
		\centering
          \resizebox{1.02\linewidth}{!}{
%
%
\definecolor{mycolor1}{rgb}{0.00000,0.44700,0.74100}%
\definecolor{mycolor2}{rgb}{0.85000,0.32500,0.09800}%
\definecolor{mycolor3}{rgb}{0.92900,0.69400,0.12500}%
\begin{tikzpicture}

\begin{axis}[%
width=1\textwidth,
scale only axis,
xmin=1,
xmax=20,
ymin=1,
ymax=1.1,
title= {$N$$=$$6$, $\Gamma$$=$$0.6 \Gamma_{\max}$},
xlabel style={font=\color{white!15!black}},
xlabel style={font=\normalsize},
xlabel={Maximum effective error variance ($\gamma_{\max}$)},
xtick = {1,5,10,15,20},
ylabel style={font=\normalsize},
ylabel style={font=\color{white!15!black}},
ylabel={Suboptimality Ratio},
axis background/.style={fill=white},
ytick = {1,1.02,1.04,1.06,1.08,1.1},
xmajorgrids,
ymajorgrids,
]
\addplot [color=mycolor1, densely dotted, line width=2.0pt]
  table[row sep=crcr]{%
1	1\\
2	1.00400879543221\\
3	1.00748554011263\\
4	1.01280036866652\\
5	1.03313748697747\\
6	1.05811640602805\\
7	1.07037039383824\\
8	1.08468427600152\\
9	1.08617943130741\\
10	1.09352320057121\\
11	1.07751356461021\\
12	1.07863381627904\\
13	1.07505178885645\\
14	1.06817002961238\\
15	1.07255112225981\\
16	1.05568106474728\\
17	1.05711987025541\\
18	1.04806720563565\\
19	1.04965103540641\\
20	1.03763833279131\\
};

\addplot [color=mycolor2, densely dashed, line width=2.0pt]
  table[row sep=crcr]{%
1	1\\
2	1\\
3	1.0004811859582\\
4	1.01127122722298\\
5	1.03094528682224\\
6	1.05275609515468\\
7	1.05537660543993\\
8	1.05463694943041\\
9	1.05102593448658\\
10	1.05675646237603\\
11	1.04158620505736\\
12	1.02886938943328\\
13	1.03240948979133\\
14	1.03396608871805\\
15	1.03843203595826\\
16	1.01482304297618\\
17	1.01142403867406\\
18	1.01234129662329\\
19	1.01826479818715\\
20	1.01048364804166\\
};

\addplot [color=mycolor3, line width=2.0pt]
  table[row sep=crcr]{%
1	1.02989954448225\\
2	1.00525979344381\\
3	1.07941354573486\\
4	1.05237613074364\\
5	1.04286306561482\\
6	1.03631367134662\\
7	1.07366260826657\\
8	1.07755803719377\\
9	1.07588968889186\\
10	1.07565751250624\\
11	1.06451828238518\\
12	1.06724956298227\\
13	1.06484282978634\\
14	1.05823635329947\\
15	1.06186062272602\\
16	1.05005508684863\\
17	1.0509669939215\\
18	1.04441039191797\\
19	1.0443653633962\\
20	1.03249686548537\\
};

\end{axis}
\end{tikzpicture}%
}
          \end{subfigure}
	\begin{subfigure}[t]{0.32\linewidth}
	\centering
          \resizebox{1\linewidth}{!}{
%
%
\definecolor{mycolor1}{rgb}{0.00000,0.44700,0.74100}%
\definecolor{mycolor2}{rgb}{0.85000,0.32500,0.09800}%
\definecolor{mycolor3}{rgb}{0.92900,0.69400,0.12500}%
\begin{tikzpicture}

\begin{axis}[%
width=1\textwidth,
scale only axis,
xmin=1,
xmax=20,
ymin=1,
ymax=1.2,
title= {$N$$=$$8$, $\Gamma$$=$$0.6 \Gamma_{\max}$},
xlabel style={font=\color{white!15!black}},
xlabel style={font=\normalsize},
xlabel={Maximum effective error variance ($\gamma_{\max}$)},
xtick = {1,5,10,15,20},
axis background/.style={fill=white},
ytick = {1,1.05,1.1,1.15,1.2},
xmajorgrids,
ymajorgrids,
]
\addplot [color=mycolor1, densely dotted, line width=2.0pt]
  table[row sep=crcr]{%
1	1\\
2	1.0004427732591\\
3	1.02085940945504\\
4	1.02309499256004\\
5	1.02318344980327\\
6	1.04183025753421\\
7	1.05298408264904\\
8	1.06565628089829\\
9	1.07399203696726\\
10	1.08285885086364\\
11	1.08393777624045\\
12	1.07635697280441\\
13	1.07545204132639\\
14	1.06974943883202\\
15	1.06744612598225\\
16	1.072957737497\\
17	1.0553035018201\\
18	1.05541071181401\\
19	1.04879689924314\\
20	1.03819016017709\\
};

\addplot [color=mycolor2, densely dashed, line width=2.0pt]
  table[row sep=crcr]{%
1	1\\
2	1.0004427732591\\
3	1.02085940945504\\
4	1.02109766120399\\
5	1.02318344980327\\
6	1.040301078629\\
7	1.05220802155254\\
8	1.06496534540781\\
9	1.0712052871458\\
10	1.07836062125634\\
11	1.07725870765283\\
12	1.0650769194378\\
13	1.06402855172647\\
14	1.05435379279648\\
15	1.04976950033503\\
16	1.05783569432286\\
17	1.03731769099752\\
18	1.03976544530732\\
19	1.03586307828363\\
20	1.02580877495495\\
};

\addplot [color=mycolor3, line width=2.0pt]
  table[row sep=crcr]{%
1	1.19333710692881\\
2	1.06013276179264\\
3	1.12493105941724\\
4	1.17265640204421\\
5	1.10410393419452\\
6	1.05550542423502\\
7	1.05141824520999\\
8	1.06682435539647\\
9	1.0502281954575\\
10	1.07971007093167\\
11	1.06236909523771\\
12	1.06115046572051\\
13	1.06564658041369\\
14	1.07055850610816\\
15	1.05795664514492\\
16	1.06391368214959\\
17	1.04788743008785\\
18	1.05023186437412\\
19	1.04316094936479\\
20	1.03471298846741\\
};

\end{axis}
\end{tikzpicture}%
}
          \end{subfigure}
	\begin{subfigure}[t]{0.32\linewidth}
	\centering
          \resizebox{1\linewidth}{!}{
%
%
\definecolor{mycolor1}{rgb}{0.00000,0.44700,0.74100}%
\definecolor{mycolor2}{rgb}{0.85000,0.32500,0.09800}%
\definecolor{mycolor3}{rgb}{0.92900,0.69400,0.12500}%
\begin{tikzpicture}

\begin{axis}[%
width=1\textwidth,
scale only axis,
xmin=1,
xmax=20,
ymin=1,
ymax=1.3,
title= {$N$$=$$10$, $\Gamma$$=$$0.6 \Gamma_{\max}$},
xlabel style={font=\color{white!15!black}},
xlabel style={font=\normalsize},
xlabel={Maximum effective error variance ($\gamma_{\max}$)},
xtick = {1,5,10,15,20},
axis background/.style={fill=white},
ytick = {1,1.05,1.1,1.15,1.2,1.25,1.3},
xmajorgrids,
ymajorgrids,
legend style={legend cell align=left, align=left, draw=white!15!black}
]
\addplot [color=mycolor1, densely dotted, line width=2.0pt]
  table[row sep=crcr]{%
1	1\\
2	1.00237393047629\\
3	1.0086008389586\\
4	1.0390048839565\\
5	1.04417266504471\\
6	1.03439029018936\\
7	1.04756790742763\\
8	1.06037845083832\\
9	1.05714853820286\\
10	1.05871241829502\\
11	1.06328389181064\\
12	1.05387216618405\\
13	1.05313648302175\\
14	1.05574342396859\\
15	1.0454801439276\\
16	1.04010245873998\\
17	1.04272785466351\\
18	1.04454662795256\\
19	1.03149225321627\\
20	1.03040870865244\\
};
\addlegendentry{Greedy}

\addplot [color=mycolor2, densely dashed, line width=2.0pt]
  table[row sep=crcr]{%
1	1\\
2	1.00093268231535\\
3	1.0086008389586\\
4	1.03759571127843\\
5	1.04417266504471\\
6	1.03439029018936\\
7	1.04756790742763\\
8	1.06037845083832\\
9	1.05714853820286\\
10	1.05871241829502\\
11	1.06328389181064\\
12	1.05387216618405\\
13	1.05313648302175\\
14	1.05574342396859\\
15	1.0454801439276\\
16	1.03921601619826\\
17	1.04272785466351\\
18	1.04454662795256\\
19	1.03149225321627\\
20	1.03040870865244\\
};
\addlegendentry{DLG}

\addplot [color=mycolor3, line width=2.0pt]
  table[row sep=crcr]{%
1	1.10512708489413\\
2	1.25289529114186\\
3	1.08686249443146\\
4	1.20150233696834\\
5	1.16256822294975\\
6	1.10531576256753\\
7	1.08857397534226\\
8	1.06939062538632\\
9	1.05629592533258\\
10	1.044632380353\\
11	1.04300059539359\\
12	1.04644248312476\\
13	1.04703539249806\\
14	1.04594872994399\\
15	1.0407480979462\\
16	1.0386356048015\\
17	1.03374997346261\\
18	1.03667390242604\\
19	1.02626265297467\\
20	1.02643263588353\\
};
\addlegendentry{DoS}

\end{axis}
\end{tikzpicture}%
}
          \end{subfigure}
	\vspace{0.5cm}\\
    \centering
	\begin{subfigure}[t]{0.34\linewidth}
	\centering
          \resizebox{1.04\linewidth}{!}{
%
%
\definecolor{mycolor1}{rgb}{0.00000,0.44700,0.74100}%
\definecolor{mycolor2}{rgb}{0.85000,0.32500,0.09800}%
\definecolor{mycolor3}{rgb}{0.92900,0.69400,0.12500}%
\begin{tikzpicture}

\begin{axis}[%
width=1\textwidth,
scale only axis,
xmin=0,
xmax=1,
xlabel style={font=\color{white!15!black}},
xlabel style={font=\normalsize},
xlabel={Normalized threshold $\beta$$=$$\Gamma/ \Gamma_{\max}$},
title= {$N$$=$$6$, $\gamma_{\max}$$=$$10$},
ymin=1,
ymax=1.1,
ylabel style={font=\normalsize},
ylabel style={font=\color{white!15!black}},
ylabel={Suboptimality Ratio},
xmajorgrids,
ymajorgrids,
axis background/.style={fill=white},
]
\addplot [color=mycolor1, densely dotted, line width=2.0pt]
  table[row sep=crcr]{%
0	1\\
0.25	1\\
0.3	1.00256157093424\\
0.35	1.05461140028971\\
0.4	1.08003187512606\\
0.45	1.05342459391274\\
0.5	1.01191422124599\\
0.55	1.09519592341763\\
0.6	1.08632337482162\\
0.65	1.03508046094967\\
0.7	1.06500906458249\\
0.75	1.0705529903942\\
0.8	1.03616919295011\\
0.85	1.00264473745082\\
0.9	1\\
1	1\\
};

\addplot [color=mycolor2, densely dashed, line width=2.0pt]
  table[row sep=crcr]{%
0	1\\
0.3	1\\
0.35	1.00122174775379\\
0.4	1.0089730055755\\
0.45	1.03016104134305\\
0.5	1.01039698472923\\
0.55	1.01188186255497\\
0.6	1.05757338417855\\
0.65	1.03305857705905\\
0.7	1.00436062576973\\
0.75	1.02606201000647\\
0.8	1.02946195193402\\
0.85	1.00264473745082\\
0.9	1\\
1	1\\
};

\addplot [color=mycolor3, line width=2.0pt]
  table[row sep=crcr]{%
0	1\\
0.05	1\\
0.1	1.00843166965221\\
0.15	1.02140716603143\\
0.2	1.05127591253975\\
0.25	1.02332617462754\\
0.3	1.06876426951394\\
0.35	1.045278094926\\
0.4	1.06817371659088\\
0.45	1.05651516485242\\
0.5	1.05211736171158\\
0.55	1.08441055536726\\
0.6	1.07716742428939\\
0.65	1.01914685407505\\
0.7	1.06267677686467\\
0.75	1.07143172469563\\
0.8	1.02685109447023\\
0.85	1.00077147977226\\
0.9	1\\
1	1\\
};

\end{axis}
\end{tikzpicture}%
} 	\end{subfigure}
	\begin{subfigure}[t]{0.32\linewidth}
	\centering
          \resizebox{1\linewidth}{!}{
%
%
\definecolor{mycolor1}{rgb}{0.00000,0.44700,0.74100}%
\definecolor{mycolor2}{rgb}{0.85000,0.32500,0.09800}%
\definecolor{mycolor3}{rgb}{0.92900,0.69400,0.12500}%
\begin{tikzpicture}

\begin{axis}[%
width=1\textwidth,
scale only axis,
xmin=0,
xmax=1,
ymin=1,
xlabel style={font=\color{white!15!black}},
xlabel style={font=\normalsize},
xlabel={Normalized threshold $\beta$$=$$\Gamma/ \Gamma_{\max}$},
title= {$N$$=$$8$, $\gamma_{\max}$$=$$10$},
ymax=1.4,
axis background/.style={fill=white},
xmajorgrids,
ymajorgrids,
]
\addplot [color=mycolor1, densely dotted, line width=2.0pt]
  table[row sep=crcr]{%
0	1\\
0.15	1\\
0.2	1.00224308091253\\
0.25	1.04071920328184\\
0.3	1.08763994232882\\
0.35	1.04508303788258\\
0.4	1.0822918417366\\
0.45	1.1163281151446\\
0.5	1.03709412502104\\
0.55	1.11141846213716\\
0.6	1.07798548106811\\
0.65	1.06302532577132\\
0.7	1.09324853447708\\
0.75	1.03092902752247\\
0.8	1.07253103194897\\
0.85	1.03864497434334\\
0.9	1.00107909740729\\
0.95	1\\
1	1\\
};

\addplot [color=mycolor2, densely dashed, line width=2.0pt]
  table[row sep=crcr]{%
0	1\\
0.2	1\\
0.25	1.00095665678423\\
0.3	1.01304057190397\\
0.35	1.03708240124343\\
0.4	1.01551143457121\\
0.45	1.08276806383622\\
0.5	1.03700723155613\\
0.55	1.06087140725993\\
0.6	1.07366871897557\\
0.65	1.01459098371381\\
0.7	1.07819825062222\\
0.75	1.03092902752247\\
0.8	1.03429715391428\\
0.85	1.03524732116412\\
0.9	1.00107909740729\\
0.95	1\\
1	1\\
};

\addplot [color=mycolor3, line width=2.0pt]
  table[row sep=crcr]{%
0	1\\
0.05	1\\
0.1	1.17400977133675\\
0.15	1.16869876763602\\
0.2	1.3266529120384\\
0.25	1.36977038295231\\
0.3	1.1599262563267\\
0.35	1.10279519043858\\
0.4	1.1393214645457\\
0.45	1.14151724335503\\
0.5	1.04694791314637\\
0.55	1.12031236135187\\
0.6	1.06182734370302\\
0.65	1.05425673321681\\
0.7	1.07944211857048\\
0.75	1.01625611141584\\
0.8	1.07020798760616\\
0.85	1.02992014595449\\
0.9	1.00029141231761\\
0.95	1\\
1	1\\
};

\end{axis}
\end{tikzpicture}%
} 	\end{subfigure}
	\begin{subfigure}[t]{0.32\linewidth}
	\centering
          \resizebox{1\linewidth}{!}{
%
%
\definecolor{mycolor1}{rgb}{0.00000,0.44700,0.74100}%
\definecolor{mycolor2}{rgb}{0.85000,0.32500,0.09800}%
\definecolor{mycolor3}{rgb}{0.92900,0.69400,0.12500}%
\begin{tikzpicture}

\begin{axis}[%
width=1\textwidth,
scale only axis,
xmin=0,
xmax=1,
title= {$N$$=$$10$, $\gamma_{\max}$$=$$10$},
xlabel style={font=\color{white!15!black}},
xlabel style={font=\normalsize},
xlabel={Normalized threshold $\beta$$=$$\Gamma/ \Gamma_{\max}$},
ymin=1,
ymax=1.7,
axis background/.style={fill=white},
xmajorgrids,
ymajorgrids,
legend style={legend cell align=left, align=left, draw=white!15!black}
]
\addplot [color=mycolor1, densely dotted, line width=2.0pt]
  table[row sep=crcr]{%
0	1\\
0.15	1\\
0.2	1.04960511610233\\
0.25	1.07542291271844\\
0.3	1.02254547058476\\
0.35	1.12780649844342\\
0.4	1.04836451265282\\
0.45	1.10159746303307\\
0.5	1.07097241144815\\
0.55	1.09259869319523\\
0.6	1.06722009440799\\
0.65	1.08381918961989\\
0.7	1.05564970628963\\
0.75	1.06946349732655\\
0.8	1.03935832128884\\
0.85	1.05464691419774\\
0.9	1.01773341259382\\
0.95	1\\
1	1\\
};
\addlegendentry{Greedy}

\addplot [color=mycolor2, densely dashed, line width=2.0pt]
  table[row sep=crcr]{%
0	1\\
0.15	1\\
0.2	1.00338316985753\\
0.25	1.03371057154639\\
0.3	1.01264105702154\\
0.35	1.10330248550824\\
0.4	1.04836451265282\\
0.45	1.08782467755884\\
0.5	1.07097241144815\\
0.55	1.07964600383214\\
0.6	1.06722009440799\\
0.65	1.07465034218696\\
0.7	1.05564970628963\\
0.75	1.06224361662353\\
0.8	1.03935832128884\\
0.85	1.04997523627685\\
0.9	1.01773341259382\\
0.95	1\\
1	1\\
};
\addlegendentry{DLG}

\addplot [color=mycolor3, line width=2.0pt]
  table[row sep=crcr]{%
0	1\\
0.05	1.05765451210553\\
0.1	1.3477907158273\\
0.15	1.25750896691841\\
0.2	1.59589499301334\\
0.25	1.1315867067961\\
0.3	1.24505653267431\\
0.35	1.15137081950024\\
0.4	1.04440003026336\\
0.45	1.14896539273903\\
0.5	1.07064367835796\\
0.55	1.08065126518067\\
0.6	1.0468504818921\\
0.65	1.07025727858972\\
0.7	1.03922682821415\\
0.75	1.07213288443684\\
0.8	1.02737181694199\\
0.85	1.04813710641811\\
0.9	1.01286792277821\\
0.95	1\\
1	1\\
};
\addlegendentry{DoS}

\end{axis}
\end{tikzpicture}%
} 	\end{subfigure}
    \caption{Suboptimality ratios (SRs) of the proposed algorithms averaged over 100 randomly generated subset selection problem instances. (Top) SRs when the total number of agents is $N$$\in$$\{6,8,10\}$ and the effective localization error variances $\{\gamma_i: i\in [N]\}$ are generated randomly from the interval $(0,\gamma_{\max})$. (Bottom) SRs when the total number of agents is $N$$\in$$\{6,8,10\}$ and the expected beamforming gain threshold is $\Gamma$$=$$\beta \Gamma_{\max}$.  }
    \label{exp_var}
    \endminipage
\end{figure*}
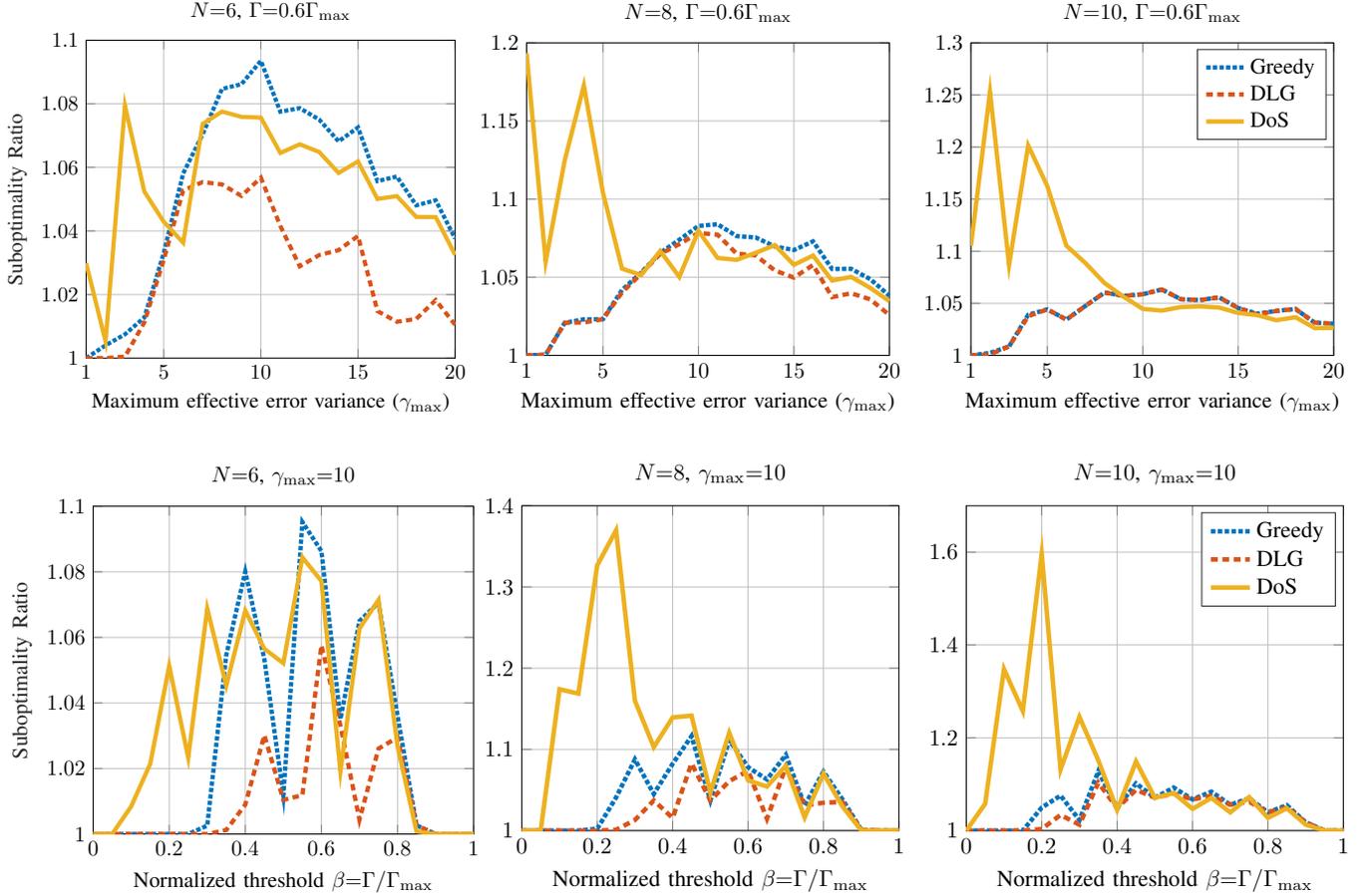

In the first set of experiments, we investigate the relationship between the algorithms' SR, the total number $N$ of agents, and the bound $\gamma_{\max}$ on the agents' effective localization error variances. For a given $N$ and $\gamma_{\max}$, a problem instance consists of $\{\gamma_i$$:$ $i$ $\in$$[N]\}$ where each $\gamma_i$ is uniformly randomly selected from the interval $(0,\gamma_{\max})$. We set the expected beamforming gain threshold as $\Gamma$$=$$0.6\Gamma_{\max}$ to allow the algorithms to output subsets of different sizes if it is optimal to do so. Recall that the DoS algorithm has only local optimality guarantees. Hence, the SR of the algorithm's output depends on the initialization of the SSP. Accordingly, for each problem instance, we run the DoS algorithm 10 times using different initializations and report the performance of the best output. Finally, we set $\lambda_0$$=$$1$ and $\alpha$$=$$2$.

For each $N$$\in$$\{6,8,10\}$ and each $\gamma_{\max}$$\in$$\{1,2,\ldots,20\}$, we generate 100 problem instances and illustrate the average SRs of all algorithms in Figure \ref{exp_var} (top). As can be seen from the figure, all algorithms show near-optimal performance (SR$\leq 1.3$) for all ($N$,$\gamma_{max}$) pairs. Recall from Theorem \ref{main_thm_1} that, when $\gamma_{\max}$$\leq$$0.83$, both the Greedy and DLG algorithms are guaranteed to have SR=$1$. Moreover, the DLG algorithm is always guaranteed to have smaller SR than the Greedy algorithm. The results shown in Figure \ref{exp_var} (top) empirically witness these theoretical guarantees. Moreover, as can be seen from the figure, both the Greedy and DLG algorithms perform well (SR $\leq 1.1$) even when the sufficient optimality condition, $\gamma_{\max}$$\leq$$0.83$, is violated. The DoS algorithm shows comparable performance to that of the Greedy and DLG algorithms when $\gamma_{\max}$$\geq$$10$. However, for small effective localization error variances, the Greedy and DLG algorithms perform significantly better than the DoS algorithm. Finally, note that the SR of the DoS algorithm increases with increasing total number $N$ of agents in general. On the other hand, the SR of the Greedy and DLG algorithms, in general, remain at the same level despite the increasing total number of agents.

In the second set of experiments, we investigate the relationship between the algorithms' SRs, the total number $N$ of agents, and the normalized threshold $\beta$$=$$\Gamma/\Gamma_{\max}$. For given $N$ and $\beta$, a problem instance consists of $\{\gamma_i : i$$\in$$[N]\}$ where each $\gamma_i$ is selected uniformly randomly from the interval $(0,10)$, i.e., $\gamma_{\max}$$=$$10$. Finally, we set $\lambda_0$$=$$1$ and $\alpha$$=$$2$, and run the DoS algorithm with 10 random initializations.

For each $N$$\in$$\{4,6,8\}$ and each $\beta$$\in$$\{0.1,0.2,\ldots,1\}$, we generate 100 problem instances. The average SRs of the algorithms over the generated instances are shown in Figure \ref{exp_var} (bottom). As can be seen from the figure, all algorithms have average SR less than $1.6$ for each $(N,\beta)$ pair. Recall from Theorem \ref{main_thm_1} that, for instances in which the threshold $\Gamma$ can be attained using two agents, the Greedy and DLG algorithms have $SR$$=$$1$. For small $\beta$ values, we observe that the Greedy and DLG algorithms achieve $SR$$=$$1$ since, in most problem instances, the threshold is attained by using two agents. Similar to the first set of experiments (Figure \ref{exp_var} (top)), we observe that the SR of the DoS algorithm increases with increasing $N$ in general. Moreover, the performance of the DoS algorithm, in general, improves with increasing $\beta$ values.

The empirical performance evaluation of the proposed algorithms on small-scale instances show that all three algorithms, Greedy, DLG, and DoS, achieve near-optimal performance (SR$\leq 1.6$) for a range of $N$, $\beta$, and $\gamma_{\max}$ values. Although the Greedy and DLG algorithms have theoretical optimality guarantees only for small $\gamma_{\max}$ and $\beta$ values, they perform considerably well (SR$\leq 1.1$) even for large $\gamma_{\max}$ and $\beta$ values. On the other hand, although the local optimality guarantee of the DoS algorithm is independent of the problem parameters, the performance of the algorithm is, in general, comparable (SR $\leq 1.1$) to that of the Greedy and DLG algorithms only for large $\gamma_{\max}$ and $\beta$ values.

\begin{figure*}[t!]
    \centering
    \minipage[t]{1\linewidth}
	\begin{subfigure}[t]{0.32\linewidth}
          \resizebox{1.01\linewidth}{!}{
%
%
\definecolor{mycolor1}{rgb}{0.00000,0.44700,0.74100}%
\definecolor{mycolor2}{rgb}{0.85000,0.32500,0.09800}%
\definecolor{mycolor3}{rgb}{0.92900,0.69400,0.12500}%
\definecolor{mycolor4}{rgb}{0.49400,0.18400,0.55600}%
\begin{tikzpicture}

\begin{axis}[%
width=1\textwidth,
at={(1.011in,0.642in)},
scale only axis,
xmin=0,
xmax=1,
ymin=0,
ytick= {0.2, 0.4, 0.6, 0.8, 1},
xlabel style={font=\color{white!15!black}},
xlabel={Normalized threshold $\beta$},
ymax=1.05,
axis background/.style={fill=white},
ylabel style={font=\color{white!15!black}},
ylabel={Normalized variance $\kappa$},
xmajorgrids,
ymajorgrids,
legend style={at= {(0.6,0.8)}, draw=white!15!black}
]
\addplot [color=mycolor1, densely dotted, line width=2.0pt]
  table[row sep=crcr]{%
0	0\\
0.02	0.000136002974748806\\
0.0600000000000001	0.000948411983385133\\
0.1	0.0031216646345773\\
0.12	0.00459597981718085\\
0.14	0.0067967656818082\\
0.16	0.00853284059910853\\
0.18	0.0114130619503034\\
0.22	0.018314497839194\\
0.24	0.0218807801504561\\
0.26	0.0249611369806748\\
0.28	0.0325639377997282\\
0.3	0.0363852448877613\\
0.32	0.043778638891812\\
0.34	0.0490789195779102\\
0.36	0.057389672608551\\
0.38	0.062233581942291\\
0.4	0.0750817879657049\\
0.42	0.0805634479403925\\
0.44	0.090694406863171\\
0.46	0.0992495849097865\\
0.48	0.112574944182569\\
0.5	0.124264150577498\\
0.52	0.138645258862354\\
0.54	0.15111394210893\\
0.56	0.169045419398488\\
0.58	0.179796511545269\\
0.6	0.204102556382235\\
0.62	0.223470334305385\\
0.64	0.246602717268646\\
0.66	0.2671057231143\\
0.68	0.296962271806645\\
0.7	0.321204854192815\\
0.72	0.352429771753139\\
0.76	0.416465669932457\\
0.78	0.45562263352651\\
0.8	0.492460594665966\\
0.82	0.532675979230797\\
0.84	0.573377266523425\\
0.86	0.614882779410018\\
0.88	0.663880394406276\\
0.9	0.717057656000519\\
0.94	0.832196700139769\\
0.96	0.887354309184614\\
0.98	0.958327915697559\\
1	1\\
};
\addlegendentry{Greedy}

\addplot [color=mycolor3, line width=2.0pt]
  table[row sep=crcr]{%
0	0\\
0.02	0.000264905910007274\\
0.04	0.00164564334586847\\
0.0600000000000001	0.00377351603776188\\
0.0800000000000001	0.0044776046042232\\
0.1	0.00941061694693057\\
0.12	0.0115555108443761\\
0.14	0.0168610716007858\\
0.16	0.019944611965121\\
0.18	0.0243239252932477\\
0.2	0.0305385716355313\\
0.22	0.0393865768482919\\
0.24	0.0386665862961559\\
0.26	0.0412124532966334\\
0.28	0.0516367279696188\\
0.3	0.0683239581655615\\
0.32	0.0746890232548101\\
0.34	0.0841042695415215\\
0.36	0.0903734902766931\\
0.38	0.0874288245752386\\
0.4	0.102694653708794\\
0.42	0.112848776507742\\
0.44	0.125994536058994\\
0.46	0.151020647625729\\
0.48	0.147229650131338\\
0.5	0.154548206515174\\
0.52	0.169009599601543\\
0.54	0.190338765273957\\
0.56	0.207240142884028\\
0.58	0.241694744824479\\
0.6	0.247855189119499\\
0.62	0.287878396563794\\
0.64	0.299229745889037\\
0.66	0.344584802732262\\
0.68	0.340473925013538\\
0.7	0.371362340018592\\
0.72	0.382206569602341\\
0.74	0.418433035152551\\
0.76	0.449742955888544\\
0.78	0.493292257805344\\
0.8	0.52618707557128\\
0.82	0.562877164536153\\
0.84	0.60333128732711\\
0.86	0.650022818616456\\
0.88	0.666422244420922\\
0.9	0.736111165084375\\
0.92	0.781819930133409\\
0.94	0.833324418458026\\
0.96	0.889546873953379\\
0.98	0.95783339323511\\
1	0.999999999999993\\
};
\addlegendentry{DoS}

\addplot [color=mycolor4, dashed, line width=2.0pt]
  table[row sep=crcr]{%
0	0\\
0.02	0.000119219645052349\\
0.04	0.000476890448220413\\
0.0600000000000001	0.00107298601175065\\
0.0800000000000001	0.00190755819700339\\
0.1	0.00298054526946312\\
0.12	0.00429203388704114\\
0.14	0.00584194393265469\\
0.16	0.00763025380087212\\
0.18	0.00965699305892764\\
0.2	0.0119221908702898\\
0.22	0.0144526781947634\\
0.24	0.0172586822567862\\
0.26	0.0204509227220251\\
0.28	0.0240315914081055\\
0.3	0.028087056921777\\
0.32	0.0327081618435896\\
0.34	0.0379256862474269\\
0.36	0.0436689863199551\\
0.38	0.0499646378226934\\
0.4	0.0570658280461096\\
0.42	0.0648329378600048\\
0.44	0.073384102688598\\
0.46	0.0827137528776445\\
0.48	0.0928606851593394\\
0.5	0.103958994635437\\
0.52	0.115994698064731\\
0.54	0.129274897866413\\
0.56	0.14379511163592\\
0.58	0.159730714094988\\
0.6	0.17721980843573\\
0.62	0.196228712593357\\
0.64	0.216963376326391\\
0.66	0.239717017485599\\
0.68	0.264506734210432\\
0.7	0.291303881938099\\
0.72	0.320277160524545\\
0.74	0.351679058938619\\
0.76	0.385539808642984\\
0.78	0.421746876056285\\
0.8	0.460410822187795\\
0.82	0.501776653836784\\
0.84	0.545833714145311\\
0.86	0.592774144667098\\
0.88	0.642523919698597\\
0.9	0.694985185525699\\
0.92	0.750359159662915\\
0.94	0.808751157312214\\
0.96	0.870021503470723\\
0.98	0.933761513699213\\
1	0.999999999976343\\
};
\addlegendentry{SDP}

\end{axis}
\end{tikzpicture}%
}
    \end{subfigure}
	\begin{subfigure}[t]{0.32\linewidth}
          \resizebox{\linewidth}{!}{
%
%
\definecolor{mycolor1}{rgb}{0.00000,0.44700,0.74100}%
\definecolor{mycolor2}{rgb}{0.85000,0.32500,0.09800}%
\definecolor{mycolor3}{rgb}{0.92900,0.69400,0.12500}%
\definecolor{mycolor4}{rgb}{0.49400,0.18400,0.55600}%
\begin{tikzpicture}

\begin{axis}[%
width=1\textwidth,
at={(1.011in,0.642in)},
scale only axis,
xmin=0,
xmax=1,
xlabel style={font=\color{white!15!black}},
xlabel={Normalized threshold $\beta$},
ymin=0,
ymax=41,
axis background/.style={fill=white},
ylabel style={font=\color{white!15!black}},
ylabel={Size of the selected subset $\mathcal{S}^{\star}$},
xmajorgrids,
ymajorgrids,
legend style={at= {(0.6,0.9)}, draw=white!15!black}
]
\addplot [color=mycolor1, densely dotted, line width=2.0pt]
  table[row sep=crcr]{%
0	1\\
0.0200000000000031	2\\
0.0399999999999991	2.9\\
0.0600000000000023	3.2\\
0.0799999999999983	3.8\\
0.100000000000001	4.3\\
0.119999999999997	4.7\\
0.140000000000001	5.3\\
0.159999999999997	5.6\\
0.200000000000003	6.6\\
0.240000000000002	7.4\\
0.259999999999998	7.7\\
0.280000000000001	8.4\\
0.299999999999997	8.7\\
0.32	9.2\\
0.340000000000003	9.6\\
0.359999999999999	10.1\\
0.380000000000003	10.4\\
0.399999999999999	11.1\\
0.420000000000002	11.4\\
0.439999999999998	11.9\\
0.460000000000001	12.3\\
0.479999999999997	12.9\\
0.5	13.4\\
0.520000000000003	14\\
0.539999999999999	14.5\\
0.560000000000002	15.2\\
0.579999999999998	15.6\\
0.600000000000001	16.5\\
0.619999999999997	17.2\\
0.640000000000001	18\\
0.659999999999997	18.7\\
0.68	19.7\\
0.700000000000003	20.5\\
0.759999999999998	23.5\\
0.780000000000001	24.7\\
0.799999999999997	25.8\\
0.859999999999999	29.4\\
0.880000000000003	30.8\\
0.899999999999999	32.3\\
0.939999999999998	35.5\\
0.960000000000001	37\\
0.979999999999997	38.9\\
1	40\\
};
\addlegendentry{Greedy}

\addplot [color=mycolor3, line width=2.0pt]
  table[row sep=crcr]{%
0	0\\
0.0200000000000031	2.1\\
0.0399999999999991	3.2\\
0.0600000000000023	4.2\\
0.0799999999999983	4.4\\
0.100000000000001	5.6\\
0.119999999999997	6\\
0.140000000000001	6.8\\
0.159999999999997	7.3\\
0.18	7.7\\
0.200000000000003	8.5\\
0.219999999999999	9.1\\
0.240000000000002	9.1\\
0.259999999999998	9.2\\
0.280000000000001	9.9\\
0.299999999999997	11\\
0.32	11.4\\
0.340000000000003	12\\
0.359999999999999	12.1\\
0.380000000000003	12\\
0.399999999999999	12.6\\
0.420000000000002	13.3\\
0.439999999999998	13.9\\
0.460000000000001	15.1\\
0.479999999999997	14.7\\
0.5	15\\
0.520000000000003	15.6\\
0.539999999999999	16.4\\
0.560000000000002	17.1\\
0.579999999999998	18.3\\
0.600000000000001	18.5\\
0.619999999999997	19.9\\
0.640000000000001	20.2\\
0.659999999999997	21.7\\
0.68	21.5\\
0.700000000000003	22.5\\
0.719999999999999	22.9\\
0.740000000000002	24\\
0.759999999999998	24.9\\
0.780000000000001	26.2\\
0.799999999999997	27.1\\
0.82	28.2\\
0.840000000000003	29.4\\
0.859999999999999	30.7\\
0.880000000000003	31\\
0.899999999999999	33\\
0.920000000000002	34.2\\
0.939999999999998	35.6\\
0.960000000000001	37.1\\
0.979999999999997	38.9\\
1	40\\
};
\addlegendentry{DoS}

\addplot [color=mycolor4, dashed, line width=2.0pt]
  table[row sep=crcr]{%
0	0\\
0.0200000000000031	40\\
1	40\\
};
\addlegendentry{SDP}

\end{axis}
\end{tikzpicture}%
}
	\end{subfigure}
	\begin{subfigure}[t]{0.32\linewidth}
          \resizebox{1.05\linewidth}{!}{
%
%
\definecolor{mycolor1}{rgb}{0.00000,0.44700,0.74100}%
\definecolor{mycolor2}{rgb}{0.85000,0.32500,0.09800}%
\definecolor{mycolor3}{rgb}{0.92900,0.69400,0.12500}%
\definecolor{mycolor4}{rgb}{0.49400,0.18400,0.55600}%
\begin{tikzpicture}

\begin{axis}[%
width=1\textwidth,
at={(1.011in,0.653in)},
scale only axis,
xmin=0,
xmax=1,
xlabel style={font=\color{white!15!black}},
xlabel={Normalized threshold $\beta$},
ymode=log,
ymin=1e-05,
ymax=100,
yminorticks=true,
ylabel style={font=\color{white!15!black}},
ylabel={Computation time (secs)},
axis background/.style={fill=white},
xmajorgrids,
ymajorgrids,
legend columns=2,
legend style={at={(0.722,0.589)}, legend cell align=left, align=left, draw=white!15!black}
]
\addplot [color=mycolor1, densely dotted, line width=2.0pt]
  table[row sep=crcr]{%
0	0.0001866\\
0.0199999999999996	0.0001328\\
0.04	5.46000000000001e-05\\
0.0599999999999996	5.93000000000001e-05\\
0.0800000000000001	9.39999999999999e-05\\
0.0999999999999996	2.28e-05\\
0.12	2.44e-05\\
0.14	2.54e-05\\
0.16	3.84e-05\\
0.18	2.38e-05\\
0.2	2.31e-05\\
0.22	2.51e-05\\
0.24	2.42e-05\\
0.26	2.5e-05\\
0.28	2.98e-05\\
0.3	2.68e-05\\
0.32	2.83e-05\\
0.34	2.9e-05\\
0.36	3.15e-05\\
0.38	3.2e-05\\
0.4	3.62e-05\\
0.42	3.82e-05\\
0.44	3.73e-05\\
0.46	3.86e-05\\
0.48	4.15e-05\\
0.5	4.62e-05\\
0.52	4.72e-05\\
0.54	5.04999999999999e-05\\
0.56	5.81e-05\\
0.58	5.63e-05\\
0.6	6.30000000000001e-05\\
0.62	6.83e-05\\
0.64	7.46000000000001e-05\\
0.66	8.11e-05\\
0.68	9.14000000000001e-05\\
0.7	0.0001012\\
0.72	0.0001173\\
0.74	0.0001387\\
0.76	0.000144\\
0.78	0.0001604\\
0.8	0.0001763\\
0.82	0.0002063\\
0.84	0.000226\\
0.86	0.0002678\\
0.88	0.000285\\
0.9	0.0003294\\
0.92	0.0003735\\
0.94	0.0004254\\
0.96	0.000479999999999999\\
0.98	0.0005736\\
1	0.000600199999999999\\
};
\addlegendentry{Greedy}

\addplot [color=mycolor2, densely dashed, line width=2.0pt]
  table[row sep=crcr]{%
0	0.0001481\\
0.0199999999999996	6.52000000000001e-05\\
0.04	6.21999999999999e-05\\
0.0599999999999996	9.39999999999999e-05\\
0.0800000000000001	0.0001495\\
0.0999999999999996	7.6e-05\\
0.12	0.0001157\\
0.14	0.0001344\\
0.16	0.0002062\\
0.18	0.0002556\\
0.2	0.000334\\
0.22	0.0004112\\
0.24	0.0005117\\
0.26	0.000644099999999999\\
0.28	0.0007453\\
0.3	0.0008569\\
0.32	0.0009987\\
0.34	0.0011256\\
0.36	0.0012726\\
0.38	0.0013645\\
0.4	0.0016258\\
0.42	0.0015932\\
0.44	0.001679\\
0.46	0.0017605\\
0.48	0.0018876\\
0.5	0.0019763\\
0.52	0.0020803\\
0.54	0.0021456\\
0.58	0.0023038\\
0.6	0.0023686\\
0.64	0.0025753\\
0.66	0.0026326\\
0.68	0.002726\\
0.7	0.0027907\\
0.72	0.0029267\\
0.74	0.0030234\\
0.76	0.0031522\\
0.78	0.0033232\\
0.8	0.00341469999999999\\
0.82	0.00353429999999999\\
0.84	0.0036375\\
0.86	0.0039015\\
0.88	0.0041086\\
0.9	0.0043725\\
0.92	0.0046325\\
0.94	0.0048797\\
0.96	0.0051142\\
0.98	0.0054934\\
1	0.00572949999999999\\
};
\addlegendentry{DLG}

\addplot [color=mycolor3, line width=2.0pt]
  table[row sep=crcr]{%
0	0.0027893\\
0.02	7.612689\\
0.04	8.5083104\\
0.0600000000000001	8.80068990000001\\
0.0800000000000001	9.6845991\\
0.1	10.3143354\\
0.12	10.2622987\\
0.14	11.006261\\
0.16	11.165624\\
0.18	10.8173302\\
0.2	11.7305465\\
0.22	12.1327946\\
0.24	12.242037\\
0.26	12.5271636\\
0.28	12.7602302\\
0.3	13.0893326\\
0.32	13.394304\\
0.34	13.9554654\\
0.36	13.8414669\\
0.38	13.8481873\\
0.4	14.7416487\\
0.42	14.8371759\\
0.44	15.4171764\\
0.46	15.1387464\\
0.48	15.4395067\\
0.52	15.9531413\\
0.54	15.9896602\\
0.56	15.7487245\\
0.58	16.955974\\
0.6	16.5056984\\
0.62	16.9640446\\
0.66	17.4371523\\
0.68	18.0042127\\
0.7	18.4963565\\
0.72	18.0528082\\
0.74	18.1218903\\
0.76	18.703705\\
0.78	19.3357524\\
0.8	19.2062025\\
0.82	19.5827738\\
0.84	19.8643363\\
0.86	20.4178271\\
0.88	20.8693184\\
0.9	21.0421272\\
0.92	21.8501204\\
0.94	22.8679966\\
0.96	22.8874696\\
0.98	23.3356364\\
1	24.8817628\\
};
\addlegendentry{DoS}

\addplot [color=mycolor4, dashed, line width=2.0pt]
  table[row sep=crcr]{%
0	0.3428439\\
0.02	0.290254\\
0.04	0.287079\\
0.0600000000000001	0.2787462\\
0.0800000000000001	0.291492\\
0.1	0.2750163\\
0.12	0.274255\\
0.14	0.2739343\\
0.16	0.2754476\\
0.18	0.2755269\\
0.22	0.2731302\\
0.24	0.2708782\\
0.26	0.2710837\\
0.28	0.2698064\\
0.3	0.2696194\\
0.32	0.2686459\\
0.34	0.2663542\\
0.36	0.2668394\\
0.38	0.2669637\\
0.4	0.2641031\\
0.42	0.2630676\\
0.44	0.2630358\\
0.46	0.2665479\\
0.48	0.2697314\\
0.5	0.2704908\\
0.52	0.2724434\\
0.54	0.267714\\
0.56	0.270057\\
0.58	0.2685082\\
0.6	0.266155\\
0.62	0.2703899\\
0.64	0.2728345\\
0.66	0.2736711\\
0.68	0.271357\\
0.7	0.2760764\\
0.72	0.2784936\\
0.74	0.2801412\\
0.76	0.278818\\
0.78	0.2820281\\
0.82	0.2895137\\
0.84	0.2934723\\
0.86	0.2894658\\
0.88	0.2937866\\
0.9	0.2979072\\
0.92	0.2949055\\
0.94	0.2988521\\
0.96	0.2998655\\
0.98	0.3066331\\
1	0.3121649\\
};
\addlegendentry{SDP}

\end{axis}
\end{tikzpicture}%
}
    \end{subfigure}
    \caption{Performance comparison of the proposed algorithms with an SDP-based beamformer. (Left) For a given threshold $\Gamma$$=$$\beta \Gamma_{\max}$, the normalized variances of the beamforming gains are similar for all approaches. (Middle) The proposed algorithms achieves the same performance by employing strict subsets of the agent network when possible. (Right) Greedy and DLG approaches synthesize beamformers orders of magnitudes faster than the SDP-based beamformer. }
    \label{exp_var_2}
    \endminipage
\end{figure*}
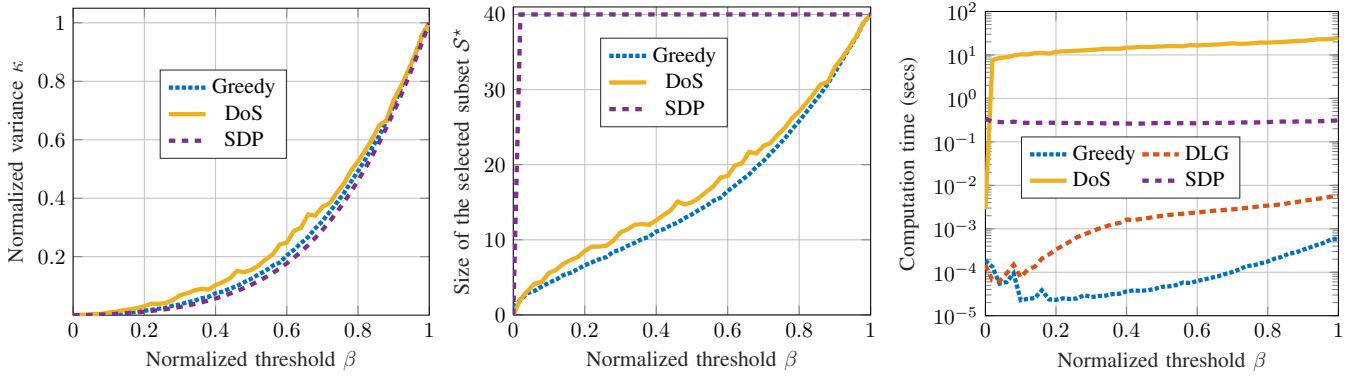

\subsection{Performance Comparison with an SDP-Based Beamformer}
We compare the performance of the proposed algorithms with a semi-definite programming-based (SDP-based) beamforming algorithm. SDP-based methods are widely used in robust beamforming literature to mitigate the degrading effects of uncertain parameters on the beam pattern \cite{gershman2010convex,kim2008robust, pascual2005robust,lorenz2005robust}. Accordingly, for comparison, we synthesize a beamforming vector ${\bf{w}}^{\star}$$\in$$\mathbb{C}^N$ where
\begin{subequations}
\begin{align}\label{SDP_1}
    {\bf{w}}^{\star}\in \arg \min_{{\bf{w}}\in \mathbb{C}^N}\ \  &\lVert {\bf{w}}\rVert_2^2\\
    \text{subject to:}\ \  &\mathbb{E}[{\bf{w}}^H{\bf{H}}{\bf{w}}]\geq \Gamma\\
   \forall i\in[N],\ \  &\lvert w_i\rvert^2 \leq 1. \label{SDP_end}
\end{align}
\end{subequations}
In \eqref{SDP_1}-\eqref{SDP_end}, the matrix ${\bf{H}}$$\in$$\mathbb{C}^{N\times N}$ is ${\bf{H}}$$=$${\bf{h}}{\bf{h}}^H$ where ${\bf{h}}^H$$=$$[h_1,h_2,\ldots,h_N]$, and ${\bf{w}}^H$$=$$[w_1,w_2,\ldots,w_N]$. The constraint in \eqref{SDP_end} ensures that $w_i$$=$$\sqrt{P}e^{j\delta_i}$ for some $P$$\leq$$1$.

A solution to the problem in \eqref{SDP_1}-\eqref{SDP_end} is a beamformer ${\bf{w}}^{\star}$ that attains the desired threshold $\Gamma$ with minimum total power while respecting the individual power constraints in \eqref{SDP_end}. It can be shown that a solution to the problem in \eqref{SDP_1}-\eqref{SDP_end} can be computed \textit{exactly} by solving an SDP \cite{gershman2010convex,luo2010semidefinite}. To synthesize the beamformer ${\bf{w}}^{\star}$, we utilized the SDP solver of the CVX toolbox \cite{cvx} with its nominal parameters. Note that the beamformer ${\bf{w}}^{\star}$ minimizes the total transmit power of the antenna array while ensuring that the expected beamforming gain exceeds the desired threshold $\Gamma$. Therefore, it represents a solution to a convex relaxation of the problem
\begin{align*}
    \min_{\mathcal{S}\subseteq[N],\delta \in \mathbb{C}^N} & \ \ \ \lvert \mathcal{S}\rvert\\ \text{subject to} &\ \ \ \mathbb{E}[G(\mathcal{S},\delta)]\geq\Gamma
\end{align*}
which is a risk-neutral version of the subset selection problem. For given ${\bf{w}}^{\star}$$=$$[w_1^{\star},w_2^{\star},\ldots, w^{\star}_N]$, we let the corresponding optimal subset be $\mathcal{S}^{\star}$$=$$\{i$$\in$$[N] : \lvert w^{\star}_i\rvert >\epsilon\}$ where $\epsilon$$=$$10^{-2}$.

We generate 100 subset selection problem instances by setting $N$$=$$40$ and selecting the error variances $\{\gamma_i : i\in [N]\}$ uniformly randomly from the interval $(0,10)$, i.e., $\gamma_{\max}$$=$$10$. For the DoS algorithm, we set $\lambda_0$$=$$1$ and $\alpha$$=$$2$, and run the algorithm with 10 random initializations. We compare the performance with respect to three metrics: the average variance of the beamforming gain, the average size of the selected subset, and the average computation time.

Figure \ref{exp_var_2} (left) shows the normalized variance of the beamforming gain, i.e., $\kappa$$=$$\mathrm{Var}(G(\mathcal{S}^{\star},\delta))/\mathrm{Var}(G([N],\hat{\delta}))$, versus the normalized threshold $\beta$$=$$\Gamma/\Gamma_{\max}$. In the figure, we do not plot the results of the DLG algorithm since it is almost exactly the same as the Greedy algorithm. As can be seen from the figure, the proposed discrete-optimization-based algorithms achieve similar performance to that of the SDP-based beamformer. The variance of the SDP-based beamformer is, in general, smaller than the variance of the proposed algorithms since the problem in \eqref{SDP_1}-\eqref{SDP_end} is a convex relaxation of the subset selection problem.

Figure \ref{exp_var_2} (middle) demonstrates the trade-off between the normalized threshold $\beta$ and the average size of the optimal subset $\mathcal{S}^{\star}$. The plot for the DLG algorithm is omitted in the figure since the average size of the subsets selected by the DLG algorithm is almost exactly the same as the Greedy algorithm. As can be seen from the figure, for $\beta$$<$$1$, the proposed algorithms employ strict subsets of the agent network $[N]$ where $N$$=$$40$. On the other hand, the SDP-based beamformer includes all the agents to beamforming for all $\beta$$>$$0$. Combined with the results shown in Figure \ref{exp_var_2} (left), this result suggests that the proposed algorithms achieve similar performance to that of the SDP-based approach using less number of agents. Hence, in a sense, the proposed algorithms improve the capabilities of the agent network as they allow the utilization of the agents that are not part of beamforming for other purposes in general.

Finally, Figure \ref{exp_var_2} (right) shows the computation times for all algorithms. As can be seen from the figure, the Greedy and DLG algorithms run orders of magnitude faster than the SDP-based beamformer. On the other hand, the DoS algorithm takes longer than the SDP-based beamformer to select a subset in general. The long computation time is partially due to the fact that we run the DoS algorithm with 10 random initializations to improve its performance. We observe in our experiments that the variance of the beamforming gain for the subset selected by the DoS algorithm decreases considerably as the number of random initializations used in the DoS algorithm increases. Therefore, there is a trade-off between the computation time of the DoS algorithm and the quality of the beamformer it synthesizes.

The empirical evaluations presented above suggests that the proposed discrete optimization-based approaches have the potential to synthesize beamformers with similar performances to that of the convex optimization-based beamformers using significantly less number of agents. Furthermore, when the Greedy and DLG algorithms are employed to synthesize beamformers, the required computation time for the synthesis can be significantly reduced with respect to SDP-based approaches.

\section{Conclusions}
We considered a network of agents that are distributed in an environment in which they locate themselves through sensor measurements and aim to transmit a message signal to a base station. Under the assumption that the agents have Gaussian localization errors, we developed three discrete optimization-based algorithms, Greedy, Double-Loop-Greedy (DLG), and Difference-of-Submodular (DoS), each of which chooses a subset of agents to optimize the quality-of-service at the base station. Specifically, the developed algorithms either globally or locally minimize the variance of the signal-to-noise ratio (SNR) received by the base station while guaranteeing that the expected SNR is above a desired threshold. We empirically showed that the proposed algorithms achieve similar performances with a convex optimization-based algorithm while using significantly less number of agents. Moreover, the Greedy and DLG algorithms run orders of magnitude faster than the convex optimization-based algorithm.

In this work, we proposed the DoS algorithm to locally minimize the variance of the received SNR when the agents' localization errors have large variances. Although the DLG algorithm achieves comparable performances to that of the convex optimization-based algorithm with less number of agents, its computational requirements may hinder its applicability to scenarios in which the size of the agent network is large. An interesting future direction might be to develop beamforming algorithms that have optimality guarantees and run fast even on large scale systems.

\bibliographystyle{IEEEtran}
\bibliography{main_conf}
\appendices\section{}
\label{appendix_A}
In this appendix, we provide proofs for all results presented in this paper. We first provide a simple technical lemma which allows us to prove the main results.

{\setlength{\parindent}{0cm}\noindent 
\begin{lemma}\label{tech_lemma} For a given set $\{x_i\in\mathbb{R} : i$$\in$$[N] \}$ of real numbers, the following equality holds:
\begin{align}
   \Bigg(\sum_{i=1}^{N}\sum_{j\neq i} x_i x_j\Bigg)^2=& 2\sum_{i=1}^N\sum_{j\neq i} x_i^2 x_j^2+4\sum_{i=1}^N\sum_{j\neq i} \sum_{\substack{k\neq i\\  k\neq j}} x_i^2 x_j x_k \nonumber \\
& +\sum_{i=1}^N\sum_{j\neq i}\sum_{\substack{k\neq i\\ k\neq j}} \sum_{\substack{l\neq i\\ l\neq j\\ l\neq k}} x_ix_jx_kx_l  \label{tech_lemma_state}
\end{align}
\end{lemma}}

\noindent \textbf{Proof of Lemma \ref{tech_lemma}:} We prove the claim by induction on $N$. 

\noindent\textit{Base case:} For the base case, i.e., $N=1$, all terms on both right and left hand sides are equal to zero since the set is a singleton. Therefore, the claim holds.

\noindent\textit{Inductive step:} Assume that the claim holds for $N$. We now show that the equality holds also for $N+1$. Using the simple formula $(a+b)^2$$=$$a^2+2ab+b^2$, we obtain
\begin{align*}
\Bigg(\sum_{i=1}^{N+1}\sum_{j\neq i} x_i x_j\Bigg)^2=&\Bigg(\sum_{i=1}^{N}\sum_{j\neq i} x_i x_j+ 2x_{N+1}\sum_{i=1}^N x_i\Bigg)^2\nonumber\\
=& A_1+ A_2 + A_3
\end{align*}
where 
\begin{align*}
    A_1=&\Bigg(\sum_{i=1}^{N}\sum_{j\neq i} x_i x_j\Bigg)^2,\\
    A_2=&4x_{N+1}\Bigg(\sum_{i=1}^N x_i\Bigg)\Bigg(\sum_{i=1}^{N}\sum_{j\neq i} x_i x_j\Bigg),\\
    A_3=&4x^2_{N+1}\Bigg(\sum_{i=1}^N x_i\Bigg)^2.
\end{align*}
By the induction hypothesis, we have
$A_1$$=$$B_1$$+$$B_2$$+$$B_3$ where 
\begin{align*}
B_1=&2\sum_{i=1}^N\sum_{j\neq i} x_i^2 x_j^2,\\  
B_2=&4\sum_{i=1}^N\sum_{j\neq i} \sum_{\substack{k\neq i\\  k\neq j}} x_i^2 x_j x_k,\\
B_3=&\sum_{i=1}^N\sum_{j\neq i}\sum_{\substack{k\neq i\\ k\neq j}} \sum_{\substack{l\neq i\\ l\neq j\nonumber\\ l\neq k}} x_i x_j x_k x_l.
\end{align*}
Using simple algebraic manipulations, we obtain 
\begin{align}
&A_3+B_1={2\sum_{i=1}^{N+1}\sum_{j\neq i} x_i^2 x_j^2}+{4x^2_{N+1}\sum_{i=1}^N\sum_{j\neq i} x_i x_j},\label{eqqqq}\\
&A_2={8x_{N+1}\sum_{i=1}^Nx^2_i\sum_{j\neq i} x_j }+4x_{N+1}\sum_{i=1}^Nx_i\sum_{j\neq i} \sum_{\substack{k\neq i\\ k \neq j}} x_j x_k.\nonumber
\end{align}
Note that the first term on the right hand side of \eqref{eqqqq} is the first term on the right hand side of \eqref{tech_lemma_state}. Let
\begin{align*}
C_1 &= B_2+4x^2_{N+1}\sum_{i=1}^N\sum_{j\neq i} x_i x_j+8x_{N+1}\sum_{i=1}^Nx^2_i\sum_{j\neq i} x_j,\\
C_2 &= B_3+4x_{N+1}\sum_{i=1}^Nx_i\sum_{j\neq i} \sum_{\substack{k\neq i\\ k \neq j}} x_j x_k.
\end{align*}
It can be shown that $C_1$ and $C_2$ are the second and third terms in the right hand side of \eqref{tech_lemma_state}. As a result, we conclude that the claim holds. $\Box$


\noindent\textbf{Proof of Proposition \ref{gaussian_power_prop22}:}
By taking the expectation of the both sides of \eqref{beamf_gain_first}, we obtain $ \mathbb{E}\Big[G(\mathcal{S},\delta)\Big]$
\begin{subequations}
\begin{align}
=&\mathbb{E}\Bigg[\sum_{i\in S}\sum_{j\in S} \cos(\Phi_i-\Phi_j)\Bigg] \\
 =& \mathbb{E}\Bigg[\lvert S \rvert +\sum_{i\in S}\sum_{j\neq i} \cos(\Phi_i-\Phi_j)\Bigg] \label{sum}
\\
 =& \lvert S \rvert+ \sum_{i\in S}\sum_{j\neq i} \mathbb{E}\Big[\cos(\Phi_i-\Phi_j)\Big] \label{linearity}\\
=&\lvert S \rvert +\sum_{i\in S}\sum_{j\neq i}\Big(\mathbb{E}[\cos(\Phi_i)\cos(\Phi_j) + \sin(\Phi_i)\sin(\Phi_j)] \Big)\label{trig}\\
=&\lvert S \rvert +\sum_{i\in S}\sum_{j\neq i}e^{-\frac{\gamma_i+\gamma_j}{2}} \Big(\cos(\theta_i)\cos(\theta_j) + \sin(\theta_i)\sin(\theta_j)\Big) \label{wrapped_proof}
\\
=&\lvert S \rvert +\sum_{i\in S}\sum_{j\neq i}\sqrt{v_i v_j}\cos(\theta_i-\theta_j). \label{expected_beam_gain_proof}
\end{align}
\end{subequations}

The equality in \eqref{sum} follows directly from the observation that $\cos(\Phi_i-\Phi_i)$$=$$\cos(0)$$=$$1$.  We obtain \eqref{linearity} by using the linearity of expectation. Equality in \eqref{trig} follows from the trigonometric identity $\cos(x-y) = \cos(x)\cos(y)+\sin(x)\sin(y)$. Under the assumption that $\Phi_i$ and $\Phi_j$ are independent for $i$$\neq$$j$, we obtain \eqref{wrapped_proof} using the characteristic function of Gaussian random variables. Finally, the equality in \eqref{expected_beam_gain_proof} follows from the definition $v_i := e^{-\gamma_i}$ and the above mentioned trigonometric identity. $\Box$ 

\noindent\textbf{Proof of Proposition \ref{gaussian_power_prop33}:} For any given $\mathcal{S}$$\subseteq$$[N]$, $ \mathbb{E}[G(\mathcal{S},\delta)]$ is maximized if and only if $(\theta_i$$-$$\theta_j)\mod 2 \pi$$=$$0$ because $\cos(x)$$\leq$$1$ for any $x$$\in$$\mathbb{R}$ and $\cos(x)$$=$$1$ if and only if $x \mod 2\pi$$=$$0$. Recalling that $\theta_i$$=$$\mathbb{E}[\eta_i]$$+$$\delta_i$, we conclude that, for any given $\mathcal{S}$$\subseteq$$[N]$, $ \mathbb{E}[G(\mathcal{S},\delta)]$ is maximized if and only if the condition stated in the proposition holds for all $i,j$$\in$$[N]$. Finally, the result $\overline{\mathcal{S}}$$=$$[N]$ follows since $\lvert S \rvert +\sum_{i\in S}\sum_{j\neq i}\sqrt{v_i v_j}$ monotonically increases with the size of $\mathcal{S}$. $\Box$

\noindent\textbf{Proof of Proposition \ref{gaussian_power_prop}:} We now derive the explicit form of the variance of the beamforming gain. Recall that $\mathrm{Var}(G(\mathcal{S},\hat{\delta}))$$=$$\mathbb{E}[G(\mathcal{S},\hat{\delta})^2]$$-$$\mathbb{E}[G(\mathcal{S},\hat{\delta})]^2$. We have
\begin{align*}
\mathbb{E}\Big[G(\mathcal{S},\hat{\delta})^2\Big]
&= \mathbb{E}\Bigg[\Bigg(\lvert S\rvert + \sum_{i\in \mathcal{S}}\sum_{j\neq i} \cos\Big(\Phi_i-\Phi_j\Big)\Bigg)^2\Bigg] \\
&=K_1+K_2+K_3
\end{align*}
where $K_1$$=$$\lvert S\rvert^2$, $K_2$$=$$2\lvert S \rvert \sum_{i\in S}\sum_{j\neq i}\sqrt{v_i v_j}$, and 
\begin{align*}
    K_3=\mathbb{E}\Bigg[\Bigg(\sum_{i\in S}\sum_{j\neq i} \cos\Big(\Phi_i-\Phi_j\Big)\Bigg)^2\Bigg].
\end{align*}
Using the trigonometric identity $2\sin(x)\cos(x)$$=$$\sin(2x)$, together with the fact that $\mathbb{E}[\sin \lambda X]$$=$$0$ for $\lambda$$\in$$\mathbb{N}$ and $X$$\sim$$\mathcal{N}(0,\sigma^2)$, it can be shown that $K_3$$=$$K_{31}$$+$$K_{32}$ where
\begin{align*}
K_{31}&=\mathbb{E}\Bigg[\Bigg(\sum_{i\in S}\sum_{j\neq i}\cos(\Phi_i)\cos(\Phi_j)\Bigg)^2\Bigg],\\
K_{32}&=\mathbb{E}\Bigg[\Bigg(\sum_{i\in S}\sum_{j\neq i}\sin(\Phi_i)\sin(\Phi_j)\Bigg)^2\Bigg].
\end{align*}

Using the characteristic function of Gaussian random variables, the identity $2\cos^2(x)-1$$=$$1-2\sin^2(x)$$=$$\cos(2x)$, and Lemma \ref{tech_lemma}, it can further be shown that
\begin{align*}
   K_{31}=&\frac{1}{2}\sum_{i\in S}\sum_{j\neq i} (1+v_i^2)(1+v_j^2)&& \\
   &+2\sum_{i\in S}\sum_{j\neq i}\sum_{\substack{k\neq i \\ k \neq j}} (1+v_i^2) \sqrt{v_j v_k} &&\nonumber \\
     &+\sum_{i\in S}\sum_{j\neq i}\sum_{\substack{k\neq i\\ k\neq j}} \sum_{\substack{l\neq i\\ l\neq j\\ l\neq k}} \sqrt{v_iv_jv_kv_l}, &&\\
 K_{32}=&\frac{1}{2}\sum_{i\in S}\sum_{j\neq i} (1-v_i^2)(1-v_j^2). &&
\end{align*}
As a result, 
\begin{align*}
    \mathbb{E}\Big[G(\mathcal{S},\hat{\delta})^2\Big]=&K_1+K_2+\sum_{i\in S}\sum_{j\neq i} (1+v_i^2 v_j^2)\\
    &+ 2\sum_{i \in S}\sum_{j\neq i}\sum_{\substack{k\neq i\\ k\neq j}}(1+v_i^2)\sqrt{v_jv_k}\nonumber \\
    &+\sum_{i\in S}\sum_{j\neq i}\sum_{\substack{k\neq i\\ k\neq j}} \sum_{\substack{l\neq i\\ l\neq j\\ l\neq k}} \sqrt{v_iv_jv_kv_l}.
\end{align*}
Similarly, for $\mathbb{E}[G(\mathcal{S},\delta)]^2$, using the characteristic function of Gaussian random variable, together with Lemma \ref{tech_lemma}, we have 
\begin{align*}
    \mathbb{E}\Big[G(\mathcal{S},\hat{\delta})\Big]^2=&\Big(\lvert S \rvert +\sum_{i\in S}\sum_{j\neq i}\sqrt{v_iv_j}\Big)^2\nonumber \\
    =&K_1+K_2+\Big(\sum_{i\in S}\sum_{j\neq i}\sqrt{v_iv_j}\Big)^2\nonumber \\
    =& K_1+K_2+2\sum_{i\in S}\sum_{j\neq i} v_iv_j\\  &+4\sum_{i \in S}\sum_{j\neq i}\sum_{\substack{k\neq i\\ k\neq j}} v_i\sqrt{v_jv_k} \\
    &\quad +\sum_{i\in S}\sum_{j\neq i}\sum_{\substack{k\neq i\\ k\neq j}} \sum_{\substack{l\neq i\\ l\neq j\\ l\neq k}} \sqrt{v_iv_jv_kv_l}.
\end{align*}
Since $\mathrm{Var}[G(\mathcal{S},\hat{\delta})]$$=$$\mathbb{E}[G(\mathcal{S},\hat{\delta})^2]$$-$$\mathbb{E}[G(\mathcal{S},\hat{\delta})]^2$, we conclude the result. $\Box$

\noindent\textbf{Proof of Proposition \ref{expected_monotone_corollary}:} Recall that $v_i$$=$$\exp(-\gamma_i)$ by definition. By taking the derivative of \eqref{expected_closed_form} with respect to $\gamma_k$ for an arbitrary $k$$\in$$\mathcal{S}$, we obtain
\begin{align*}
    \frac{\partial \mathbb{E}[G(\mathcal{S},\hat{\delta})]}{\partial \gamma_k}=-\sum_{\substack{j\in\mathcal{S}\\ j\neq k}}\exp\Bigg(-\frac{\gamma_k+\gamma_j}{2}\Bigg)\leq 0.
\end{align*}
The above inequality implies that decreasing the value of the maximum $\gamma_k$ increases the value of the expected beamforming gain. The result then follows from the definition of $V(\mathcal{S})$. $\Box$

\noindent \textbf{Proof of Theorem \ref{main_thm_1}:}
We first show that if $\mathbb{E}[G(\mathcal{S},\hat{\delta})]$$\geq$$\Gamma$ for $\mathcal{S}$$=$$\{i_1,i_2\}$, then a solution to the problem in \eqref{opt_uncertainty_1}-\eqref{opt_uncertainty_2} is a solution to the subset selection problem. Recall that $\gamma_{i_1}$$\leq$$\gamma_{i_2}$$\leq$$\ldots$$\leq$$\gamma_{i_N}$. Then, when $\mathbb{E}[G(\mathcal{S},\hat{\delta})]$$\geq$$\Gamma$ for $\mathcal{S}$$=$$\{i_1,i_2\}$, the subset $\mathcal{S}$ is a solution to the problem in \eqref{opt_uncertainty_1}-\eqref{opt_uncertainty_2}. Now, observe from \eqref{variance_closed_form} that among the subsets $\mathcal{S}'$$\subseteq$$[N]$ that satisfy $\lvert \mathcal{S}' \rvert$$=$$2$, the subset $\mathcal{S}$$=$$\{i_1,i_2\}$ is the one that minimizes $\mathrm{Var}(G(\mathcal{S}',\hat{\delta}))$. Moreover, since all the terms in the right hand side of \eqref{variance_closed_form} are nonnegative, adding a new element to a subset $\mathcal{S}'$$\subseteq$$[N]$ that satisfy $\lvert \mathcal{S}' \rvert$$=$$2$ can only increase the value of $\mathrm{Var}(G(\mathcal{S}',\hat{\delta}))$. Consequently, the subset $\mathcal{S}$$=$$\{i_1,i_2\}$ is the subset that satisfies the constraint $\mathbb{E}[G(\mathcal{S},\hat{\delta})]$$\geq$$\Gamma$ and minimizes the variance of the beamforming gain. Hence, the claim holds.

We will now show that the second condition in the statement of the theorem is also a sufficient condition for the equivalence of the problems in \eqref{opt_uncertainty_1}-\eqref{opt_uncertainty_2} and \eqref{opt_main_1}-\eqref{opt_main_2} in terms of optimal solutions.
Without loss of generality, let
\begin{align}\label{gaussian_ineq}
    0.83\geq \gamma_1\geq \gamma_2\geq \ldots \geq \gamma_N\geq 0.
\end{align} We take the derivative of $\mathrm{Var}(G(\mathcal{S},\hat{\delta}))$ with respect to $\gamma_1$ and show that the derivative is always nonnegative. Equivalently, we show that the variance decreases as the maximum effective error variance $\gamma_i$ decreases. 

Recall that $v_i$$=$$\exp(-\gamma_i)$. After some algebra, we obtain 
\begin{align}\label{derivative_var_gauss}
\frac{\partial \mathrm{Var}(G(\mathcal{S},\hat{\delta}))}{\partial \gamma_1} =&  4\sum_{j\neq 1} v_1v_j (1-v_1v_j)\nonumber\\
&+4v_1\Big(1-v_1\Big)\sum_{j\neq 1}\sum_{\substack{k\neq 1 \\ k \neq j}}\sqrt{v_jv_k}\nonumber\\
&-2\sum_{j\neq 1}(1- v_j)^2\sum_{\substack{k\neq 1 \\ k \neq j}} \sqrt{v_1v_k}.
\end{align}
Using \eqref{gaussian_ineq}, we obtain the following three inequalities which will be used to bound each term on the right hand side of \eqref{derivative_var_gauss}:
\begin{align*}
   \text{1)}\ & \sum_{\substack{j\in S \\ j\neq 1}} v_1v_j (1-v_1v_j ) \geq (\lvert S\rvert -1) v_1^2(1-v_1)\\
    \text{2)} \ & \sum_{\substack{j\in S \\ j\neq 1}}\sum_{\substack{k\neq 1 \\ k \neq j}}\sqrt{v_jv_k}\geq 2 \binom{\lvert S \rvert -1}{2} v_1\\
   \text{3)} \ &\sum_{\substack{j\in S \\ j\neq 1}}\sum_{\substack{k\neq 1 \\ k \neq j}}(1- v_1)^2 \sqrt{v_1v_k}\leq2 \binom{\lvert S \rvert -1}{2} (1-v_1)^2\sqrt{v_1}
\end{align*}
Consequently, we have
\begin{align}
    \frac{\partial \mathrm{Var}(G(\mathcal{S},\delta))}{\partial \gamma_1}\geq& 4(\lvert S\rvert -1) v_1^2(1-v_1) \nonumber \\
    & +8\binom{\lvert S \rvert -1}{2}v_1^2(1-v_1)\nonumber\\
    & -4\binom{\lvert S \rvert -1}{2}(1-v_1)^2\sqrt{v_1}.
\end{align}
By rearranging the terms and using that $\gamma_1$$\leq$$0.83$, we obtain
\begin{align*}
    \frac{\partial \mathrm{Var}(G(\mathcal{S},\hat{\delta}))}{\partial \gamma_1^2}\geq& 2(\lvert S\rvert -1)\sqrt{v_1} (1-v_1)\times \nonumber\\
    & \Bigg[2(\lvert S\rvert -1)(v_1)^{3/2}-(\lvert S \rvert-2)(1-v_1)\Bigg] \\
    \geq& 0.
\end{align*}
Finally, since the above inequality implies that the variance can be decreased by decreasing the value of $\gamma_1$, an optimal solution to the problem in \eqref{opt_uncertainty_1}-\eqref{opt_uncertainty_2} is an optimal solution to the subset selection problem when $\gamma_1$$\leq$$0.83$.  $\Box$


\noindent\textbf{Proof of Theorem \ref{supermodularity_theorem}:} We first establish the supermodularity of $\mathbb{E}[G(\mathcal{S},\hat{\delta})]$. For notational simplicity, let $\overline{G}(\mathcal{S}) $$:=$$ G(\mathcal{S}, \hat{\delta})$. For $X,Y$$\subseteq$$[N]$ such that $X$$\subseteq$$Y$, let $X'$$=$$X\cup \{e\}$ and $Y'$$=$$Y\cup\{e\}$ where $e$$\in$$[N]$$\backslash$$Y$.
We have,
\begin{align*}
    X_{\text{diff}}:=\mathbb{E}[\overline{G}(X')]-\mathbb{E}[\overline{G}(X)]=1+2\sqrt{v_e}\sum_{i\in X}\sqrt{v_i},\\
     Y_{\text{diff}}:=\mathbb{E}[\overline{G}(Y')]-\mathbb{E}[\overline{G}(Y)]=1+2\sqrt{v_e}\sum_{i\in Y}\sqrt{v_i}.
\end{align*}
Using the fact that $v_i\geq 0$ and $X$$\subseteq$$Y$, we obtain
\begin{align*}
    X_{\text{diff}}-Y_{\text{diff}}=-2\sqrt{v_e}\sum_{i\in Y\backslash X}\sqrt{v_i} \leq 0.
\end{align*}
Hence, we conclude that $\mathbb{E}[G(\mathcal{S},\hat{\delta})]$ is supermodular.

We now show the supermodularity of $\mathrm{Var}(G(\mathcal{S},\hat{\delta}))$. For $X,Y$$\subseteq$$[N]$ such that $X$$\subseteq$$Y$, let $X'$$=$$X\cup \{e\}$ and $Y'$$=$$Y\cup\{e\}$ where $e$$\in$$[N]$$\backslash$$Y$. Then, we have
\begin{align*}
    \overline{X}_{\text{diff}}:&=\mathrm{Var}\Big(\overline{G}(X')\Big)-\mathrm{Var}\Big(\overline{G}(X)\Big)\\ 
    &=2\sum_{i\in X}(1-v_e v_i)^2+2(1-v_e)^2\sum_{i\in X}\sum_{j\neq i} \sqrt{v_i v_j}\\
    &\quad + 4\sqrt{v_e}\sum_{i\in X}\sum_{j\neq i} (1-v_i)^2\sqrt{v_j},\\
     \overline{Y}_{\text{diff}}:&=\mathrm{Var}\Big(\overline{G}(Y')\Big)-\mathrm{Var}\Big(\overline{G}(Y)\Big)\\
     &=2\sum_{i\in Y}(1-v_e v_i)^2+2(1-v_e)^2\sum_{i\in Y}\sum_{j\neq i} \sqrt{v_i v_j}\\
    &\quad + 4\sqrt{v_e}\sum_{i\in Y}\sum_{j\neq i} (1-v_i)^2\sqrt{v_j}.
\end{align*}
Using the fact that $v_i\geq 0$ and $X$$\subseteq$$Y$, it is then straightforward to show that $ X_{\text{diff}}-Y_{\text{diff}}$$\leq$$0$. This concludes the proof. $\Box$




\begin{IEEEbiography}[{\includegraphics[width=1in,height=1.25in,clip,keepaspectratio]{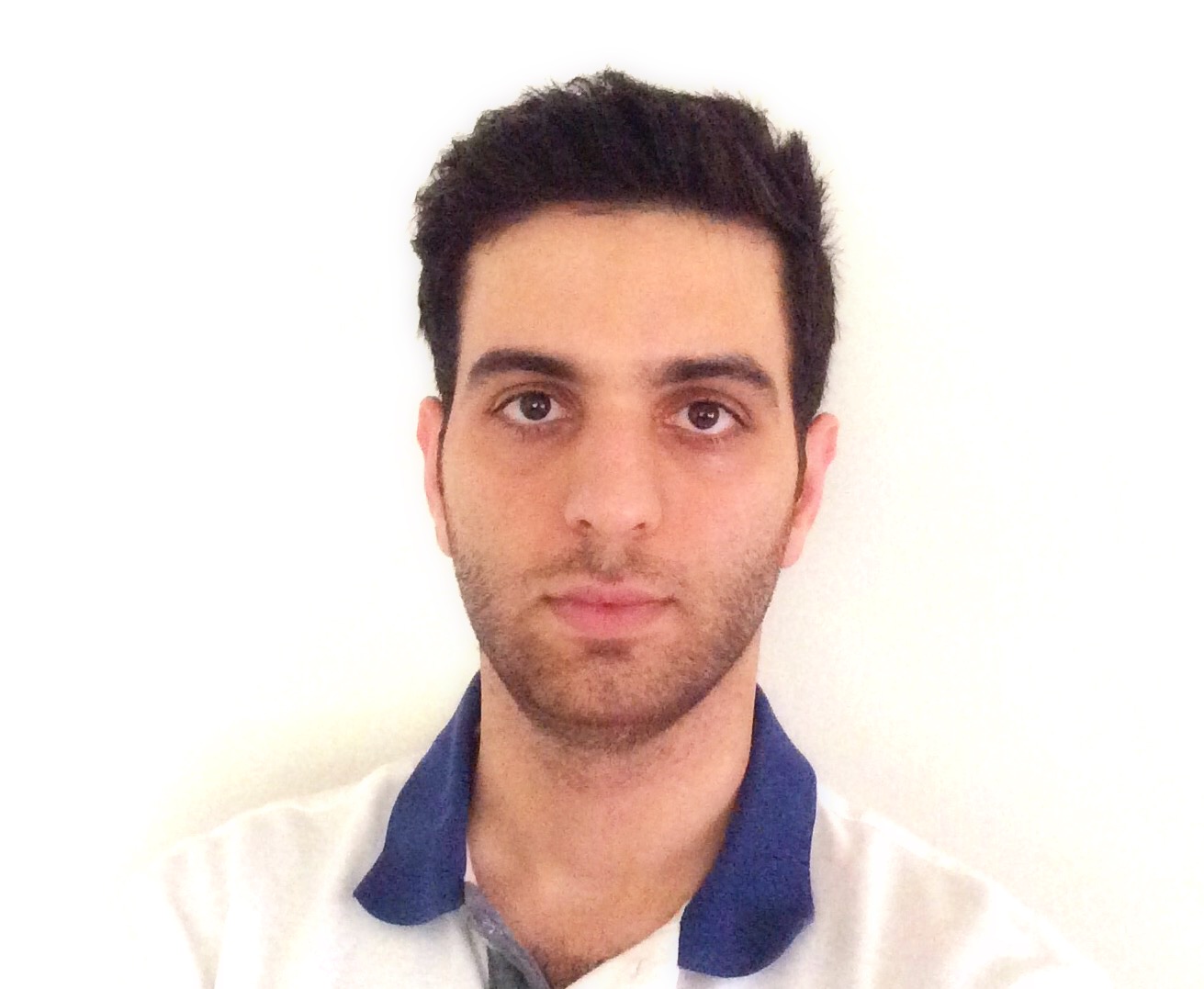}}] {Erfaun Noorani} is a Ph.D. student and a Clark Doctoral Fellow in the Department of Electrical and Computer Engineering, and a member of the Institute for Systems Research at University of Maryland College Park, College Park, United States. He received his Bachelor of Science degree in Electrical Engineering with a minor in Computer Science from Drexel University, Philadelphia, United States. His research concerns developing robust-resilient-adaptive Multi-agent Reinforcement Learning systems.
\end{IEEEbiography}
\begin{IEEEbiography}[{\includegraphics[width=1in,height=1.25in,clip,keepaspectratio]{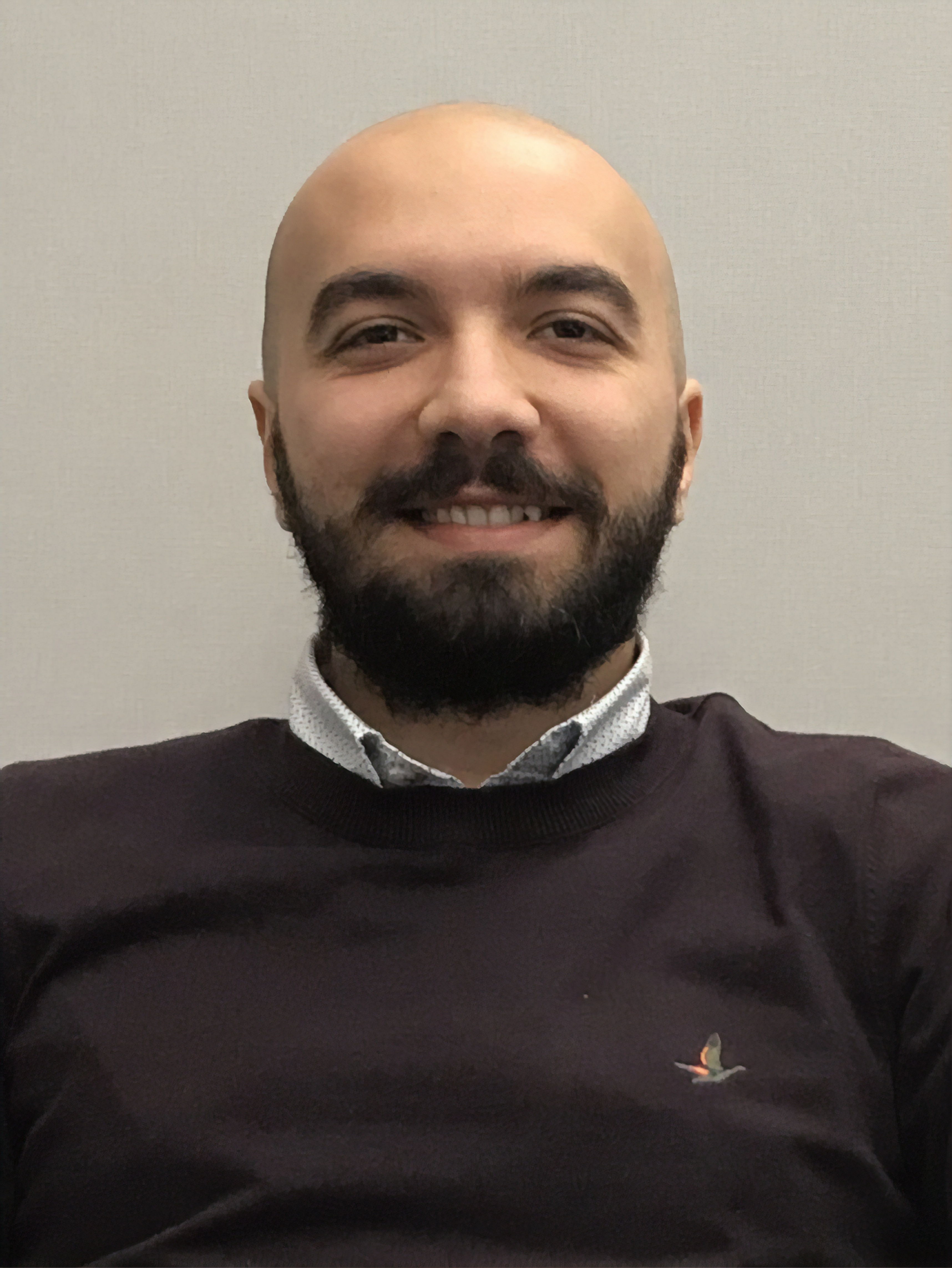}}]{Yagiz Savas} is a Ph.D. student in the Department of Aerospace Engineering at the University of Texas at Austin. He received his B.S. degree in Mechanical Engineering from Bogazici University in 2017. He is broadly interested in designing autonomous systems that operate in uncertain environments with provable safety and performance guarantees. His research focuses on developing theory and algorithms for sequential decision making under incomplete information through novel connections between controls, information theory, and formal methods.
\end{IEEEbiography}
\begin{IEEEbiography}[{\includegraphics[width=1in,height=1.25in,clip,keepaspectratio]{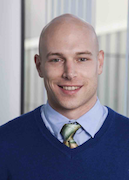}}] {Alec Koppel} is a Research Scientist at the U.S. Army Research Laboratory in the Computational and Information Sciences Directorate since September of 2017. He completed his Master’s degree in Statistics and Doctorate in Electrical and Systems Engineering, both at the University of Pennsylvania (Penn) in August of 2017. Before coming to Penn, he completed his Master’s degree in Systems Science and Mathematics and Bachelor’s Degree in Mathematics, both at Washington University in St. Louis (WashU), Missouri. He is a recipient of the 2016 UPenn ESE Dept. Award for Exceptional Service, an awardee of the Science, Mathematics, and Research for Transformation (SMART) Scholarship, a co-author of Best Paper Finalist at the 2017 IEEE Asilomar Conference on Signals, Systems, and Computers, and a finalist for the ARL Honorable Scientist Award 2019. His research interests are in optimization and machine learning. Currently, he focuses on scalable Bayesian learning, reinforcement learning, and decentralized optimization, with an emphasis on problems arising in robotics and autonomy.
\end{IEEEbiography}

\begin{IEEEbiography}[{\includegraphics[width=1in,height=1.25in,clip,keepaspectratio]{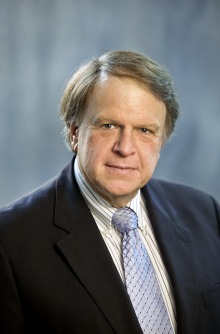}}]
{John S. Baras} is a Distinguished University Professor and holds the Lockheed Martin
Chair in Systems Engineering, with the Department of Electrical and Computer
Engineering and the Institute for Systems Research (ISR), at the University of Maryland
College Park. From 1985 to 1991, he was the Founding Director of the ISR. Since 1992,
he has been the Director of the Maryland Center for Hybrid Networks (HYNET), which
he co-founded. He received the Ph.D. degree in Applied Mathematics from Harvard
University.

He is a Fellow of IEEE (Life), SIAM, AAAS, NAI, IFAC, AMS, AIAA, Member of the National
Academy of Inventors and a Foreign Member of the Royal Swedish Academy of Engineering
Sciences. Major honors include the 1980 George Axelby Award from the IEEE Control Systems
Society, the 2006 Leonard Abraham Prize from the IEEE Communications Society, the 2017
IEEE Simon Ramo Medal, the 2017 AACC Richard E. Bellman Control Heritage Award, the
2018 AIAA Aerospace Communications Award. In 2016 he was inducted in the A. J. Clark
School of Engineering Innovation Hall of Fame. In 2018 he was awarded a Doctorate Honoris
Causa by his alma mater the National Technical University of Athens, Greece.

His research interests include systems and control, optimization, communication networks,
applied mathematics, machine learning, artificial intelligence, signal processing, robotics,
computing systems, security, trust, systems biology, healthcare systems, model-based systems
engineering.
\end{IEEEbiography}

\begin{IEEEbiography}[{\includegraphics[width=1in,height=1.25in,clip,keepaspectratio]{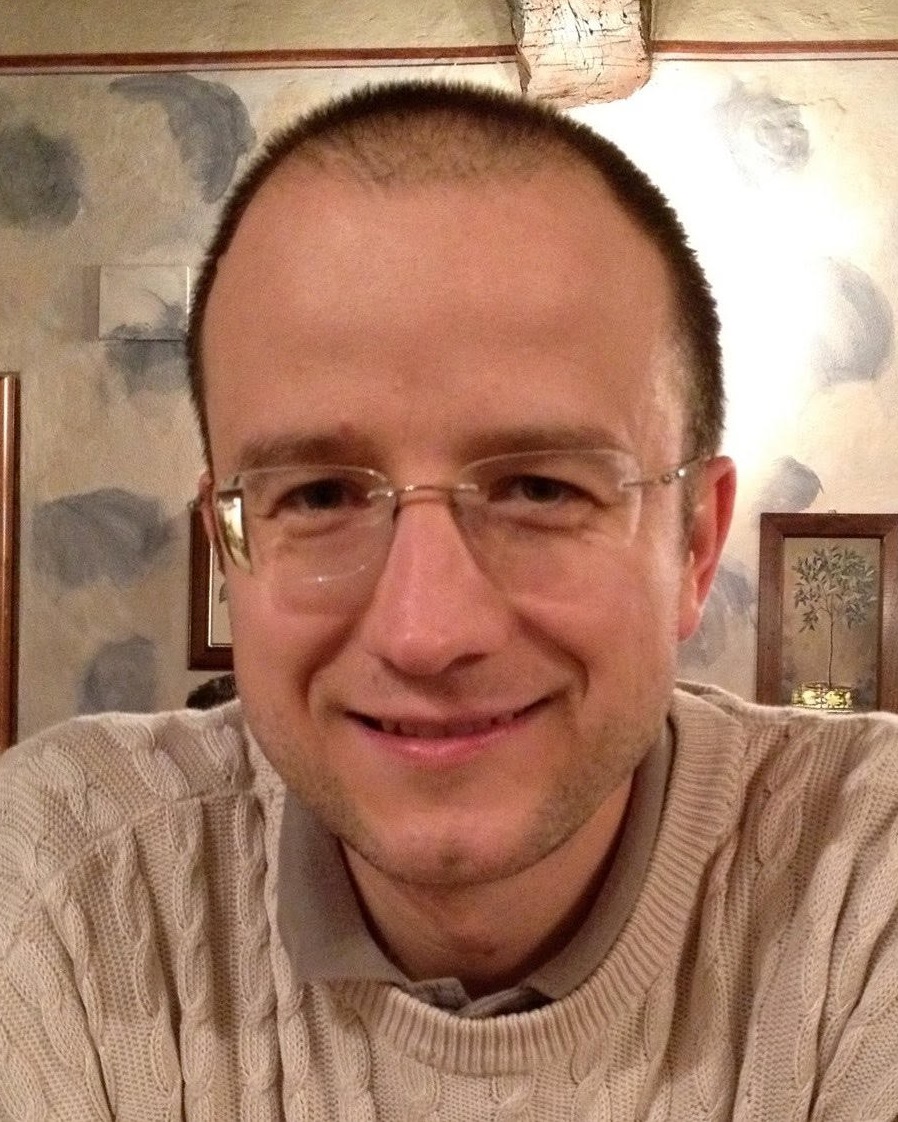}}]{Ufuk Topcu} joined the Department of Aerospace Engineering at the University of Texas at Austin as an assistant professor in Fall 2015. He received his Ph.D. degree from the University of California at Berkeley in 2008. He held research positions at the University of Pennsylvania and California Institute of Technology. His research focuses on the theoretical, algorithmic and computational aspects of design and verification of autonomous systems through novel connections between formal methods, learning theory and controls.
\end{IEEEbiography}
\begin{IEEEbiography}[{\includegraphics[width=1in,height=1.25in,clip,keepaspectratio]{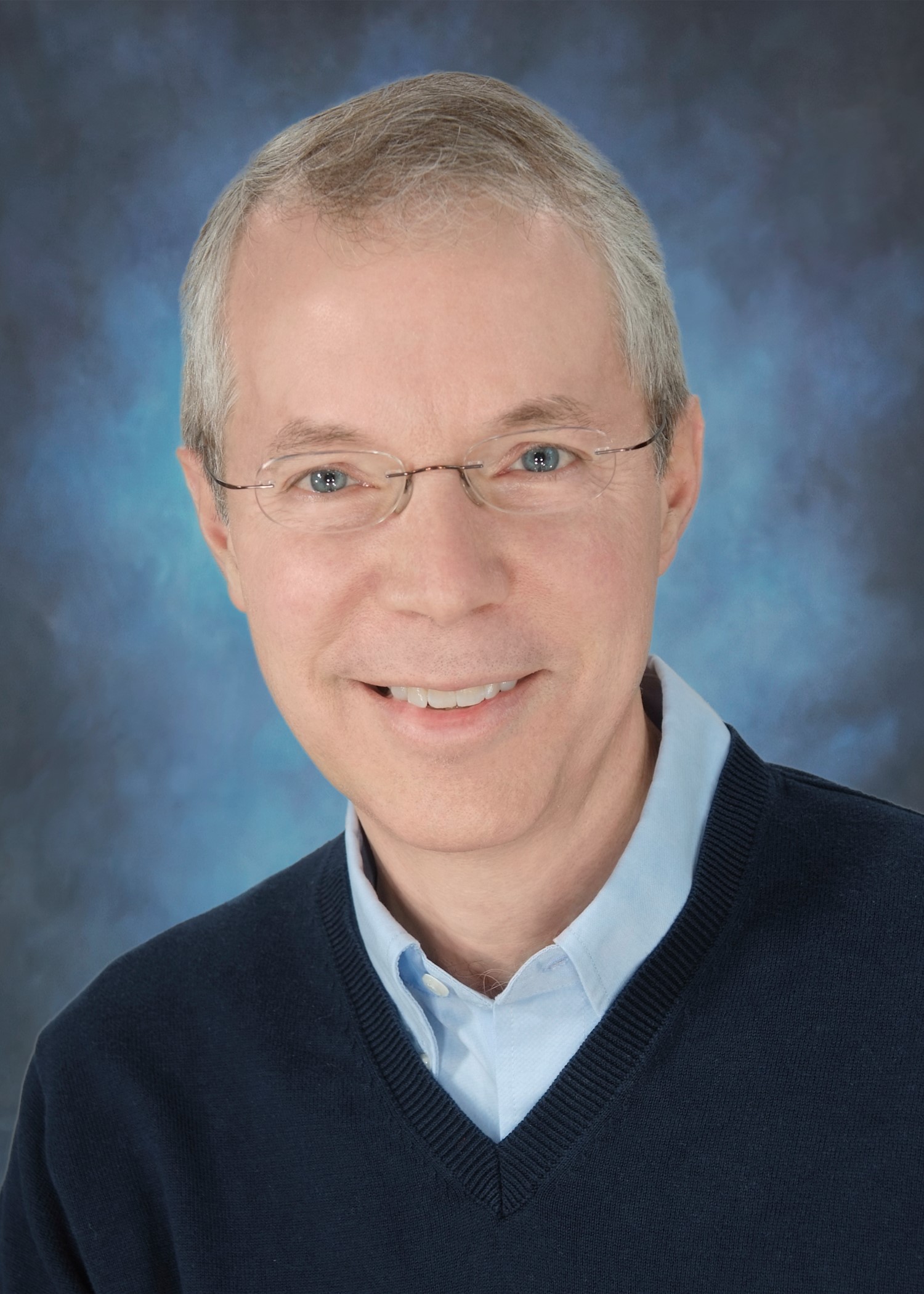}}] {Brian M. Sadler} (Life Fellow, IEEE) received the B.S. and M.S. degrees from the University of Maryland, College Park, and the PhD degree from the University of Virginia, Charlottesville, all in electrical engineering.  He is the US Army Senior Scientist for Intelligent Systems, and a Fellow of the Army Research Laboratory (ARL) in Adelphi, MD.  He has been an IEEE Communications Society Distinguished Lecturer since 2020, was an IEEE Signal Processing Society Distinguished Lecturer for 2017-2018, and general co-chair of IEEE GlobalSIP’16.  He has been an Associate Editor of the IEEE Transactions on Signal Processing, IEEE Signal Processing Letters, and EURASIP Signal Processing, and a Guest Editor for several journals including the IEEE JSTSP, the IEEE JSAC, IEEE Transactions on Robotics, the IEEE SP Magazine, Autonomous Robots, and the International Journal of Robotics Research. He received Best Paper Awards from the IEEE Signal Processing Society in 2006 and 2010, several ARL and Army R\&D awards, and a 2008 Outstanding Invention of the Year Award from the University of Maryland.  He has more than 400 publications in these areas with over 17,000 citations and h-index of 56. His research interests include information science, and networked collaborative autonomous intelligent systems.
\end{IEEEbiography}

\end{document}